\documentclass[12pt]{article}

\usepackage{amsmath}
\usepackage{graphicx,psfrag,epsf}
\usepackage{enumerate}
\usepackage{natbib}
\usepackage{url} % not crucial - just used below for the URL 

\usepackage{times}
\usepackage{balance}
\usepackage{epsfig}
\usepackage{amsthm}
\usepackage{amssymb}
\usepackage{setspace}
\usepackage{multirow}
\usepackage{subfigure}
\RequirePackage[colorlinks,citecolor=blue,urlcolor=blue]{hyperref}
\usepackage{breakurl}
\usepackage{upgreek}
\usepackage{bbm}
\usepackage{doi}
\usepackage[margin=1in]{geometry}

\newcommand{\bfs}{{\bf s}}

\newcommand{\bftheta}{{\boldsymbol \theta}}
\newcommand{\bfbeta}{{\boldsymbol \beta}}
\newcommand{\bfalpha}{{\boldsymbol \alpha}}

\def\bSig\mathbf{\Sigma}

\DeclareMathOperator{\logit}{logit}

\DeclareMathOperator{\Cov}{Cov}

\DeclareMathOperator{\FDR}{FDR}
\DeclareMathOperator{\FDP}{FDP}
\DeclareMathOperator{\FNR}{FNR}
\DeclareMathOperator{\FD}{FD}
\DeclareMathOperator{\FN}{FN}
\DeclareMathOperator{\pFDR}{pFDR}
\DeclareMathOperator{\mFDR}{mFDR}
\DeclareMathOperator{\mFNR}{mFNR}

\newcommand{\blind}{1}

\begin{document}

\def\spacingset#1{\renewcommand{\baselinestretch}%
{#1}\small\normalsize} \spacingset{1}

\if1\blind
{
  \title{\bf Spatially-Dependent Multiple Testing Under Model Misspecification, with Application to Detection of Anthropogenic Influence on Extreme Climate Events
}
  \author{Mark D. Risser \thanks{
The authors gratefully acknowledge \textit{the Office of Science, Office of Biological and Environmental Research of the U.S. Department of Energy.}}\hspace{.2cm}\\
    Climate and Ecosystem Sciences Division, Lawrence 
              Berkeley National Laboratory\\
    and \\
    Christopher J. Paciorek \\
    Department of Statistics, University of California, Berkeley\\
    and \\
    D\'aith\'i A. Stone \\
    Computational Research Division, Lawrence Berkeley National
		  Laboratory  }
  \maketitle
} \fi

\if0\blind
{
  \title{\bf Spatially-Dependent Multiple Testing Under Model Misspecification, with Application to Detection of Anthropogenic Influence on Extreme Climate Events
}
  \maketitle
} \fi

\bigskip
\begin{abstract}
The Weather Risk Attribution Forecast (WRAF) is a forecasting tool that uses output from global climate models to make simultaneous attribution statements about whether and how greenhouse gas emissions have contributed to extreme weather across the globe. However, in conducting a large number of simultaneous hypothesis tests, the WRAF is prone to identifying false ``discoveries.'' A common technique for addressing this multiple testing problem is to adjust the procedure in a way that controls the proportion of true null hypotheses that are incorrectly rejected, or the false discovery rate (FDR). Unfortunately, generic FDR procedures suffer from low power when the hypotheses are dependent, and techniques designed to account for dependence are sensitive to misspecification of the underlying statistical model. In this paper, we develop a Bayesian decision theoretic approach for dependent multiple testing and a nonparametric hierarchical statistical model that flexibly controls false discovery and is robust to model misspecification. We illustrate the robustness of our procedure to model error with a simulation study, using a framework that accounts for generic spatial dependence and allows the practitioner to flexibly specify the decision criteria. Finally, we apply our procedure to several seasonal forecasts and discuss implementation for the WRAF workflow.
\end{abstract}

\noindent%

\vfill

\newpage
\spacingset{1.45} % DON'T change the spacing!
%============================================
% SECTION 1: Introduction
%============================================
\section{Introduction} \label{section1}

Event attribution (EA) is a field of study that seeks to understand and describe the influence of greenhouse gas emissions and other human activities on extreme weather \citep{StottPA_AllenMR_etalii_2010,NAS_2016}.  The increasing interest in this field arises from the realization that a major fraction of past, current, and future climate impacts and climate change-related impacts result from the occurrence of extreme weather \citep{ArentDJ_TolRSJ_etalii_2014,SmithKR_WoodwardA_etalii_2014}.  Risk-based EA studies quantify the effect of greenhouse gas (GHG) emissions and other anthropogenic factors on weather by comparing two climate scenarios: a factual real-world scenario (the ``world as it is'') and a counterfactual, non-anthropogenic world (the ``world as it might have been'').  Then, using a probabilistic framework (\citealp{AllenM_2003, StoneDA_AllenMR_2005a, HansenG_AuffhammerM_SolowAR_2014}), a risk-based EA study compares the probabilities of pre-defined unusual weather in the two scenarios and estimates how much more or less likely extreme events are in the anthropogenically-influenced world than they would have been otherwise (note: here and throughout we mean ``risk'' in the sense of epidemiological or relative risk, not statistical risk).  Typically, the probabilities for each of these scenarios are estimated from simulations of climate models.

Risk-based EA studies can either be targeted or systematic in their approach.  Targeted studies examine one event (or a small number of events), studying in detail the meteorological mechanisms involved in the event and how the anthropogenic influence is transmitted through them, and are generally reactive in the sense that they are only conducted for an event that has actually occurred \citep[e.g.][]{StottPA_StoneDA_AllenMR_2004,PallP_AinaT_etalii_2011}.  Systematic studies cover a much larger number of events using an identical method for all events, but the rigidity of a single experimental design means that some events are not amenable to investigation \citep{AngelilO_StoneD_etalii_2016}.  An advantage of the systematic approach is that it does not necessarily depend on the occurrence of events, with it being possible to instead perform the analyses on a pre-defined list of events.  This is the approach taken by the Weather Risk Attribution Forecast (WRAF, \if1\blind{\url{http://climate.web.runbox.net/wraf}}\fi \if0\blind{[blinded]}\fi).  In order to have EA information available in ``real-time,'' the WRAF performs analyses one month in advance using a pre-defined list of 232 extreme weather events, comprising an unusually hot, cold, wet, and/or dry month over each of 58 regions \citep{AngelilO_StoneDA_etalii_2014}.  In the upcoming new version, the number of regions will be increased by a factor of about four (see, e.g., Figure~\ref{hotJan_wetMarch}; D. Stone, ``A hierarchical collection of political/economic regions for analysis of climate extremes'', submitted).  Data for both the factual and counterfactual scenarios come from climate model simulations.

Formally, the forecast involves estimating the probability of a pre-defined extreme event for both climate scenarios in each of the regions. For region $i = 1, \dots, M$ (in the upcoming version of the WRAF, $M=237$), the forecast uses the ratio of scenario-specific probabilities $p_{Fi}$ (for the factual scenario) and $p_{Ci}$ (for the counterfactual scenario) or ``risk ratio'' $RR_i = p_{Fi}/p_{Ci}$ to formally test for changes in the probability of an extreme month. In other words, a collection of statistical tests are conducted that have null hypotheses of the form
\begin{equation} \label{nullHyp}
H_i: RR_i \leq c, \hskip3ex i = 1, \dots, M,
\end{equation}
where, for example, $c=1$ if we are interested in determining whether anthropogenic influence has resulted in an increase in the event probability. Ultimately, we wish to separately test collections of hypotheses like (\ref{nullHyp}) for extreme temperature (both hot and cold) and precipitation (both wet and dry).

Of course, when the number of tests $M$ is large, a classical testing procedure is prone to identifying false ``discoveries,'' or incorrectly rejecting null hypotheses (commonly referred to as Type I errors). As such, the testing procedure is often adjusted, attempting to control the false discovery rate (FDR), which is the proportion of true null hypotheses that are incorrectly rejected. Since the data arise from physical climate models, it is anticipated that the hypotheses might be dependent: in other words, there is likely strong dependence within each spatial field of probabilities. This dependence might arise from the spatial proximity of the regions (i.e., strong dependence between $p_{Fi}$ and $p_{Fj}$ for adjacent regions $i$ and $j$) but also from potentially un-specified long-range teleconnections (in which two probabilities $p_{Fi}$ and $p_{Fj}$ might be highly correlated even if regions $i$ and $j$ are far apart) that are common for atmospheric climate variables considered over the globe (see, e.g., \citealp{CressieWikle}). Unfortunately, while classical FDR procedures (\citealp{BenjaminiHochberg1995}; see Section \ref{section2}) are theoretically valid for positively correlated hypotheses (\citealp{BenjaminiYekutieli2001}), they are also known to suffer from low power when the test statistics from each test are not independent (e.g., see \citealp{SunCai2009}). And, while the literature contains a number of methods for applying FDR procedures under dependence, the methods are outlined for specific underlying probability models and are sensitive to improper specification of this model (\citealp{Sun2015}). 

In this paper, we develop an approach to the multiple testing problem for spatially-dependent hypotheses in a systematic and decision-theoretic framework. Focusing on procedures that account for dependence among tests, we provide an overview of the diverse literature on false discovery control, including traditional methods and both Frequentist and Bayesian decision-theoretic approaches. The framework we use, originally introduced by \citealp{Muller2004}, allows the practitioner to flexibly specify the decision criteria for false discovery control, and we explore practical comparison of various FDR procedures and decision criteria that can be used when an empirical estimate of the correlation among tests is available. Furthermore, we introduce a robust yet practical modeling framework for addressing spatial dependence among hypotheses, address sensitivity of the decision rule's performance to statistical model misspecification, and demonstrate the robustness of our modeling framework for FDR control. While the methodology is designed specifically for the hypothesis testing setting of the WRAF, our framework is useful for a broader set of problems involving multiple testing over a spatial domain, particularly in the case where an empirical correlation estimate is available, which is often the case for climate science scenarios. In this context, the methods outlined in this paper could be used for general hierarchical Bayesian models beyond just considering the probability of extremes or the risk ratio. 

A reader familiar with the climate science literature will be aware of the concept of statistical field significance (\citealp{LivezeyChen1983}), which is an alternative multiple testing approach that seeks to evaluate the collective significance of a set of statistics. While field significance techniques are well-established in climate science, we instead seek to control FDR following the arguments outlined in \cite{Ventura2004}, the most important of which is that field significance provides no specific information about which individual tests are significant.
Interestingly, the idea of FDR-control is growing in popularity among climate scientists, as evidenced by a recent paper by \cite{Wilks2016}.
Finally, note that if one is only concerned with a real-time attribution statement for a single region in advance, then the multiple testing framework presented here is not required.

The paper proceeds as follows. In Section \ref{section2}, we introduce a decision theoretic framework for FDR control and present a systematic and flexible Bayesian approach to the problem. In Section \ref{section3}, we introduce our nonparametric Bayesian framework for modeling the factual and counterfactual probabilities and extreme ratios, while in Section \ref{section4} we conduct a simulation study to assess the sensitivity of the Bayesian FDR procedure to misspecification of the statistical model and identify a data-driven approach that robustly controls the FDR. In Section \ref{section5}, we apply the method to a real data set to be used for the WRAF; Section \ref{section6} concludes the paper.

%============================================
% SECTION 2: decision theory
%============================================
\section{Decision theoretic approaches for false discovery control} \label{section2}

The WRAF is generated based on monthly simulations of the Community Atmospheric Model version 5.1 (CAM5.1; see Section \ref{section42} for more details). Temperature and precipitation from the CAM5.1 climate model ensembles are aggregated monthly for each region, for both the factual ($F$) and counterfactual ($C$) scenarios. A climate model ensemble is a set of climate model runs such that each ensemble member has the same boundary conditions (for example, atmospheric chemistry or sea ice concentrations) but stochastically perturbed initial conditions. Denote the resulting collection of random variables $\{ Y_{kil}: i = 1, \dots, M; k\in\{F, C\}; l = 1, \dots, n_{\text{ens}} \}$, where $Y$ generically represents either average monthly temperature or total monthly precipitation and $n_{\text{ens}}$ is the ensemble size (or number of replicates). Formally, for an extreme event type (e.g., cold months, wet months) in region $i = 1, \dots, M$, define random variables
\begin{equation} \label{zDefs}
Z_{Fi} = \sum_{l=1}^{n_{\text{ens}}} I(Y_{Fil} > y_i), \hskip3ex
Z_{Ci} = \sum_{l=1}^{n_{\text{ens}}} I(Y_{Cil} > y_i)
\end{equation}
(for hot and wet extremes; replace ``$>$'' with ``$<$'' for cold and dry extremes), where the extreme event is defined in terms of a region-specific threshold $y_i$ (e.g., exceeding a monthly average temperature threshold of 290 K; note that the event definition is common across scenarios). Define scenario-specific data ${\bf Z} = \{ (Z_{Fi}, Z_{Ci}) : i = 1, \dots, M\}$; these binomial random variables can be used to estimate the event probabilities in each scenario $\{ (p_{Fi}, p_{Ci}) \}$ and establish evidence regarding the null hypotheses $\{{H}_i: i = 1, \dots, M\}$ from (\ref{nullHyp}).

For each null hypothesis, define a corresponding collection of unknown parameters that represent the true state for each hypothesis:
\[
\theta_i = \left\{ \begin{array}{cl} 0 & \text{if the true state of hypothesis $i$ is null} \\ 1 & \text{if the true state of hypothesis $i$ is non-null} \end{array} \right. \hskip2ex i = 1,\dots, M.
\]
The testing problem involves generating a decision rule $\boldsymbol{\delta} \equiv \boldsymbol{\delta}({\bf Z}) = \{ \delta_i : i = 1, \dots, M\}$, such that 
\[
\delta_i = \left\{ \begin{array}{cl} 0 & \text{if hypothesis $i$ is classified as null} \\ 1 & \text{if hypothesis $i$ is classified as non-null} \end{array} \right. \hskip2ex i = 1,\dots, M.
\]
In addition to specifying the form of the decision rule (often based on a test statistic, $P$-value, etc.), an underlying probability model must be specified in order to estimate the decision rule. The false discovery proportion ($\FDP$) is defined as
$\FDP = \big[\sum_{i=1}^M (1-\theta_i)\delta_i \big]/\big[1 \vee \sum_{i=1}^M \delta_i\big]$.
Note that the $\FDP$ is simply a function of unknown parameters ($\theta_i$) and random variables ($\delta_i$), and is hence fundamentally neither Frequentist nor Bayesian.

For a full summary of classical, model-specific approaches to the multiple testing problem, we refer the interested reader to Appendix A in the Supplemental Materials. The original FDR procedure given by \cite{BenjaminiHochberg1995} (henceforth BH) controls the (frequentist) FDR, defined as the expected FDP, i.e., $\FDR \equiv \mathsf{E}(\FDP)$, where the expectation is taken over repeated experiments. Their remarkably simple procedure ensures that $\FDR \leq \alpha$; the proof in \cite{BenjaminiHochberg1995} is established for independent test statistics and any configuration of false null hypotheses. One alternative to BH is the adaptive FDR procedure (\citealp{Benjamini2000}, \citealp{Genovese2002}). Other alternatives to BH are based on a random mixture model formulation of the multiple testing problem, where the $\theta_i$ are Bernoulli random variables % with $P(\theta_i = 0) = \pi_0$, 
and 
$
(Z_{Fi}, Z_{Ci}) | \theta_i \sim \theta_i F_0 + (1-\theta_i) F_1
$,
where $F_0$ and $F_1$ are the null and alternative distributions, respectively. Using this framework, procedures were developed to control either the positive FDR $\pFDR = \mathsf{E}\big(\FDP \big| \sum_{i=1}^M \delta_i > 0 \big)$ (\citealp{storey2003}), the marginal FDR $\mFDR = { \mathsf{E}\big(\sum_{i=1}^M (1-\theta_i)\delta_i\big) }/{  \mathsf{E}\big(\sum_{i=1}^M \delta_i\big) }$ (\citealp{storey2003}), and Bayesian approaches to the problem using local FDR (\citealp{Efron2001}; \citealp{Efron2004}) and the $q$-value (\citealp{storey2003}). Yet another alternative approach uses a weighted classification approach, wherein the decision rule $\boldsymbol{\delta}$ is constructed by minimizing the classification risk $\mathsf{E}[L_\lambda(\bftheta, \boldsymbol{\delta})]$, where the loss function is
\begin{equation} \label{FreqLoss}
L_\lambda(\bftheta, \boldsymbol{\delta}) = \frac{1}{M} \sum_{i=1}^M \Big\{ \lambda (1-\theta_i)\delta_i + \theta_i(1-\delta_i)   \Big\};
\end{equation}
here, $\lambda>0$ is the loss attached to a false positive error (relative to a false negative error). 

Unfortunately, proofs for the optimality of all of these procedures rely on the notion of independent hypotheses, and the optimality is called into question when the hypotheses are instead dependent. The decision rules of \cite{BenjaminiHochberg1995}, \cite{Benjamini2000}, \cite{Efron2001}, and \cite{SunCai2007} are ``simple,'' meaning that $\delta_i$ is a function only of the data corresponding to hypothesis $i$. It is easy to imagine that in the case of correlated hypotheses, compound decision rules (i.e., decision rules $\boldsymbol{\delta}$ such that $\delta_i$ depends on data corresponding to the other hypotheses) are preferred in that they might be able to identify non-nulls with a smaller signal by pooling information across tests. As a result, \cite{SunCai2009} extend the compound decision framework for multiple testing in the presence of dependence, specifically when the unknown $\theta_i$ arising from a hidden Markov model (HMM). Two recent papers by \cite{Sun2015} and \cite{Shu2015} extend this work further to provide similar results for spatial random fields and multi-dimensional Markov random fields (MRFs), respectively. However, proofs for the optimality of these procedures are model-specific; furthermore, \cite{Sun2015} also find that ``the precision of [their] testing procedure shows some sensitivity to model misspecification.''

In order to move away from the classical model-specific procedures, we are motivated to consider fully Bayesian approaches to the multiple testing problem, first presented by \cite{Newton2004}, \cite{Muller2004}, and \cite{Muller2006}. Whereas the Frequentist FDR is defined as an expectation over repeated experiments, \cite{Muller2004} defined a Bayesian FDR $\overline{\FDR} \equiv \mathsf{E}(\FDP | {\bf Z}) = \int \FDP dp(\bftheta|{\bf Z})$ (i.e., the posterior expected $\FDP$), where the expectation is with respect to the posterior distribution of the unknown states conditional on the data. Conditioning on the data and marginalizing with respect to $\bftheta$, \cite{Muller2004} showed that
$\overline{\FDR} =  \big[\sum_{i=1}^M \delta_i \pi_i \big]/\big[1 \vee \sum_{i=1}^M \delta_i \big]$,
where $\pi_i = P(\theta_i = 0 | {\bf Z})$ is the posterior probability that the $i$th hypothesis is null. A similar expression can be obtained for the Bayesian FNR,
$\overline{\FNR} = \big[{\sum_{i=1}^M (1- \delta_i)(1-\pi_i) }\big]/\big[1 \vee ({M - \sum_{i=1}^M \delta_i })\big]$,
as well as count versions $\overline{\FD} = \sum_{i=1}^M \delta_i \pi_i$ and $\overline{\FN} = \sum_{i=1}^M (1- \delta_i)(1-\pi_i)$. 

A Bayesian decision criteria that is similar in nature to the Frequentist approaches (e.g., \citealp{SunCai2007} and \citealp{Sun2015}) is to minimize the $\overline{\FNR}$ subject to the constraint that $\overline{\FDR} \leq \alpha$. \cite{Muller2004} showed that the optimal decision rule for this criteria is $\delta_i^* = I(\pi_i < t^*_\alpha)$, where the threshold depends on the desired $\alpha$. Interestingly, this decision rule can be written like the decision rule in \cite{Sun2015}: after ranking the $\pi_i$ such that $\pi_{(1)} < \pi_{(2)} < \cdots < \pi_{(M)}$, find
\begin{equation} \label{BayesR1}
r_1 = \max \left\{ j: \frac{1}{j} \sum_{i=1}^j \pi_{(i)} \leq \alpha \right\};
\end{equation}
then $t^*_\alpha = \pi_{(r_1 + 1)}$, so that we reject $H_{(1)}, \dots, H_{(r_1)}$. The difference between the decision rule in \cite{Sun2015} and (\ref{BayesR1}) is that the former involves a probability conditional on the hyperparameters while the latter involves a probability that marginalizes over the hyperparameters. In other words, the fully Bayesian posterior probability $\pi_i$ is almost the same as the oracle statistic in \cite{Sun2015}, but accounts for uncertainty in the hyperparameters. % $\boldsymbol{\xi}$. 
(Note, however, that while the oracle statistic in \citealp{Sun2015} is derived using a Frequentist criteria, it is calculated using a Bayesian framework and coincides exactly with (\ref{BayesR1}). Their simulation study verifies that this strategy controls the Frequentist FDR.)

The optimality of (\ref{BayesR1}) for controlling $\overline{\FDR} \leq \alpha$ is true for ``any probability model with non-zero prior probability for both the null and alternative hypotheses'' (\citealp{Muller2006}), which is quite powerful in light of the extensive work to develop model-specific oracle procedures in the Frequentist setting (e.g., \citealp{SunCai2007, SunCai2009, Sun2015, Shu2015}). Of course, the Bayesian FDR ($\overline{\FDR} \equiv \mathsf{E}(\FDP | {\bf Z})$) is not the same as the Frequentist FDR ($\FDR \equiv \mathsf{E}(\FDP)$), but \cite{Muller2004} and \cite{Muller2006} showed that controlling the Bayesian FDR implies Frequentist FDR control when tests are independent. Unfortunately, this is not necessarily true for dependent hypotheses (\citealp{Pacifico2004}; \citealp{Guindani2009}), although the relationship between the decision rule in \cite{Sun2015} and (\ref{BayesR1}) suggests a similarity between the two approaches.

A benefit of the decision-theoretic framework is that classification errors can be controlled in a variety of ways, beyond just the rate of false discoveries. In addition to the decision criteria that controls the Bayesian FDR introduced in the previous paragraph by minimizing a posterior expected loss (henceforth $R_1$), \cite{Muller2004} defined two other decision criteria. The first (denoted $R_2$) is similar to the classification risk for (\ref{FreqLoss}):
\begin{equation*}\label{R2}
R_2(\boldsymbol{\delta}, {\bf z}) = \lambda_1 \overline{\FD} + \overline{\FN} = \sum_{i=1}^M \big\{ \lambda_2\delta_i\pi_i + (1-\delta_i)(1-\pi_i) \big\}.
\end{equation*}
This criteria minimizes the {number} of false negatives and false discoveries, where $\lambda_2$ represents the cost for a false discovery relative to a false negative. Like $R_1$, \cite{Muller2004} showed that the optimal decision rule for $R_2$ is a threshold rule, i.e., $\delta_i^* = I(\pi_i < t^*_\lambda)$, where the optimal threshold is $t^*_\lambda = 1/(\lambda_2 + 1)$. The second (denoted $R_3$) is similar in nature to $R_1$, although instead of controlling the \textit{rate} of false discoveries we control the \textit{number} of false discoveries, i.e., $R_3$ minimizes the $\overline{\FN}$, subject to $\overline{\FD} \leq \gamma$. The optimal rule is again a threshold rule, now $\delta_i^* = I(\pi_i < t^*_\gamma)$, and we can write the optimal threshold as a step-up procedure: find 
\begin{equation} \label{BayesL3}
r_3 = \max \left\{ j: \sum_{i=1}^j \pi_{(i)} \leq \gamma \right\}
\end{equation}
and set $t^*_\gamma = \pi_{(r_3 + 1)}$ so that we reject $H_{(1)}, \dots, H_{(r_3)}$. Note that by definition, $R_2$ and $R_3$ do not specifically control the false discovery rate. However, given that the optimal decision rule for both criteria is a threshold rule (like $R_1$), they do imply FDR control at some level determined in an indirect way via $\lambda_2$ and $\gamma$.

\begin{figure}[!t]
\begin{center}
\includegraphics[trim={45 15 45 25mm}, clip, width = 0.9\textwidth]{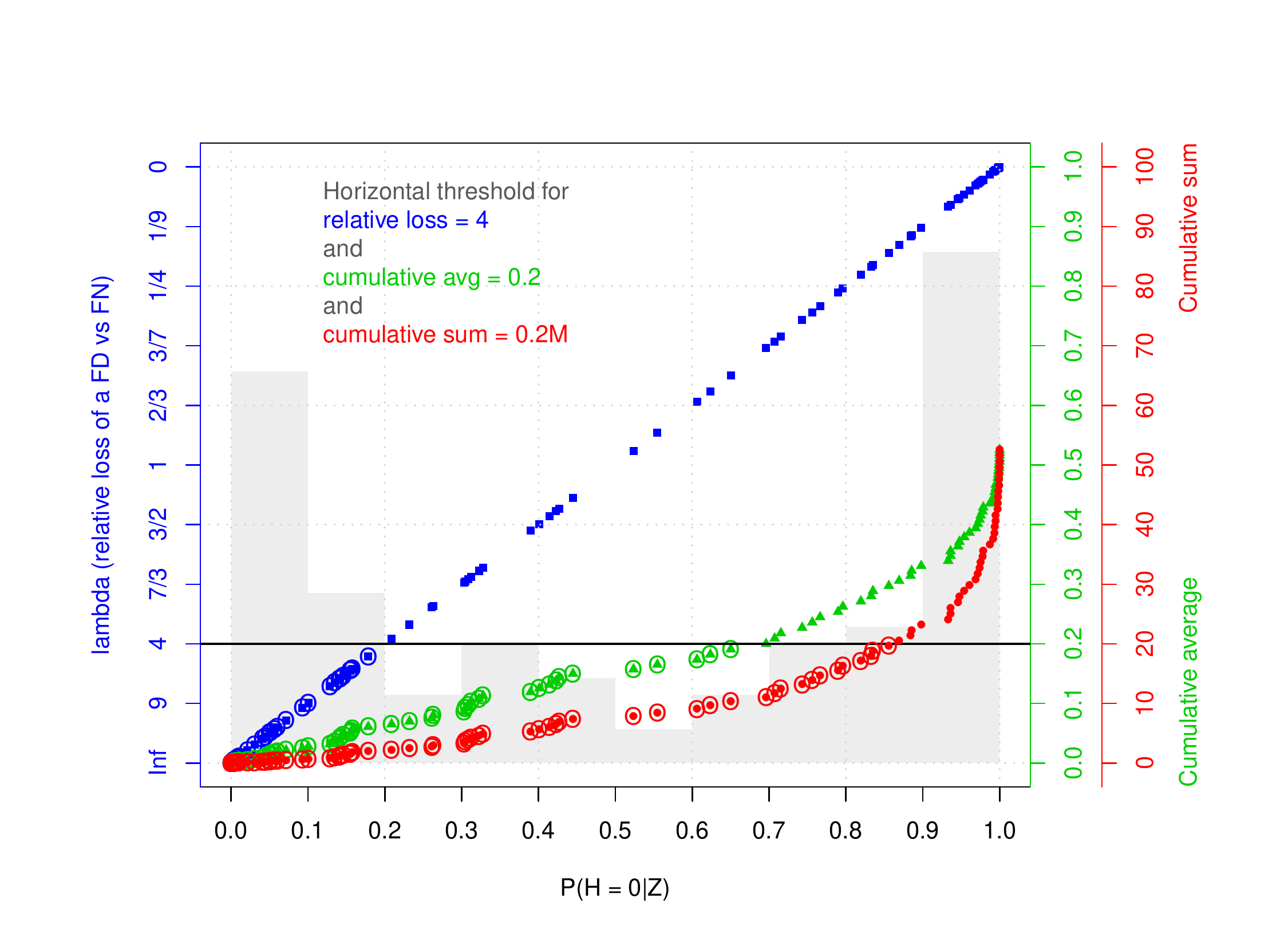}
\caption{A comparison of the various decision criteria, for a bimodal distribution of $M=100$ artificial posterior probabilities. The triangular points are plotted on the scale of $R_1$; the square points are plotted on the scale of $R_2$; the circular points are plotted on the scale of $R_3$. The horizontal threshold line illustrates the cutoff for all three decision criteria: $R_1$, where we want to make sure that fewer than 20\% of our discoveries are false; $R_2$ (which thresholds the raw probabilities), when we have specified a false discovery to be 4 times more costly than a false negative; and $R_3$, where we want to make sure that we have fewer than 20 total false discoveries.}
\label{compare_lossFcns}
\end{center}
\end{figure}

With all of these tools at our disposal, which should we use? On one hand, the three different decision criteria $R_1$, $R_2$, and $R_3$ allow the decision maker to choose a criteria based on their application of interest and what feels most natural. On the other hand, the criteria do not yield equivalent inference, even if the thresholds are ``equivalent.'' To illustrate this point, consider Figure \ref{compare_lossFcns}, which simultaneously visualizes the three criteria by plotting artificial posterior probabilities $\pi_i = P(\theta_i = 0 | {\bf Z})$ for $M=100$ tests along with the threshold statistics corresponding to $R_1$, $R_2$, and $R_3$. The $x$-axis corresponds to the posterior probabilities $\pi_i$, the light gray histogram in the background shows the distribution of the $\pi_i$, and each $y$-axis corresponds to one of the decision criteria. First, the $y$-axis on the left side of the plot displays the threshold quantity for $R_2$ (where $\pi = 1/(\lambda_2+1) \leftrightarrow \lambda_2 = 1/\pi - 1$, in blue). This axis can be thought of as the minimum $\lambda_2$ value for which a given $\pi_i$ would lead to rejection. The $y$-axes on the right show the threshold quantities for $R_1$ (the cumulative average of the $\pi_{(i)}$, in green) and $R_3$ (the cumulative sum of the $\pi_{(i)}$, in red). 

Figure \ref{compare_lossFcns} shows ``equivalent'' horizontal thresholds at $\alpha = 0.2$ for $R_1$ (meaning that we want to control $\overline{\FDR}$ at 20\%), $\lambda_2 = 4$ for $R_2$ (meaning that we specify a false discovery to be 4 times as costly as a false negative), and $\gamma = 0.2*100 = 20$ threshold for $R_3$ (meaning that we want to ensure we have fewer than 20 false discoveries). The $\pi_i$ values for rejected null hypotheses are circled. The main point of Figure \ref{compare_lossFcns} is to show how threshold values from the decision criteria relate to each other. First, if we are willing to think of the $R_1$ and $R_2$ cutoffs as equivalent (i.e., that controlling $\overline{\FDR}$ at 20\% is equivalent to a false discovery being 4 times as costly as a false negative), then we can see that $R_2$ is more conservative than $R_1$. This is true in general: $R_2$ thresholds the raw posterior probabilities $\pi_i$, while $R_1$ thresholds the cumulative average. Also, note that while a statistical model can use information from all regions to estimate the individual posterior probabilities (see Section \ref{section3}), if one uses the $R_2$ criteria then the distribution of the $\pi_i$ is unimportant: all posterior probabilities less than the threshold are classified as rejections, regardless of how they are distributed over $(0,1)$. Alternatively, $R_1$ (and $R_3$) considers the \textit{cumulative} posterior probabilities when deciding the classification rule: e.g., if there are many posterior probabilities near zero, then tests with posterior probabilities much larger than $\alpha$ can still be rejected (in Figure \ref{compare_lossFcns}, note that a test with $P(H=0|{\bf Z}) \approx 0.65$ is rejected).

Similarly, if we are willing to think of the $R_1$ and $R_3$ cutoffs as equivalent (i.e., $\alpha = \gamma/M$), then we can see that $R_1$ is more conservative than $R_3$ (again, this is true in general). However, while equating the thresholds for $R_1$ and $R_2$ is reasonable, it is much more difficult to equate the thresholds for $R_1$ and $R_3$; therefore, it might not make sense to compare $R_1$ and $R_3$. The reason for this difference is that $R_1$ considers a \textit{rate} of false discoveries, while $R_3$ considers a count: as such, the total number of discoveries or rejections is very important. For example, out of 100 tests, setting out to control the $\overline{\FDR}$ at 20\% (using $R_1$) means that if 10 tests are rejected, then having 2 of those 10 \textit{rejections} be incorrect is acceptable. This is quite different than being happy with 20 false discoveries out of 100 \textit{tests} (which is the corresponding statement for $R_3$).

Two other distributions are shown in Appendix B (in the supplementary materials), comparing the decision criteria for $\{\pi_i\}$ clustered near zero (Figure B.1) and clustered near one (Figure B.2). These figures reiterate the fact that the distribution of the $\{\pi_i\}$ is important for $R_1$ and $R_3$. When the $\pi_i$ are clustered near zero (as in Figure \ref{compare_lossFcns}), both $R_1$ and $R_3$ are quite aggressive and yield qualitatively similar results, rejecting tests for which the posterior probability of the null is large (i.e., tests where $\pi_i \approx 0.65$). Alternatively, when the $\pi_i$ are clustered near one, $R_1$ is quite conservative and rejects only a few hypotheses, while $R_3$ is still quite liberal and rejects many hypotheses. As will be seen later, in Section \ref{section4}, $R_3$ is always non-conservative: using this decision criteria will always result in rejecting \textit{at least} $\left\lfloor \gamma \right\rfloor$ tests, even when all $\pi_i = 1$.

In conclusion, we reiterate that the choice of decision criteria for a specific application depends on the criteria that feels most natural for the decision maker: indeed, this is one reason that the decision-theoretic approach is so helpful. In light of the differences in $R_1$, $R_2$, and $R_3$, our simulation study (see Section \ref{section4}) will apply each of these decision rules and summarize the performance of each in terms of their target criteria (i.e., the realized false discovery rate, loss, and false discovery count, respectively).

%============================================
% SECTION 3: nonparametric Bayes
%============================================
\section{A robust nonparametric Bayesian model with sparsity for extreme ratios} \label{section3}

A natural statistical model for the random variables ${\bf Z} = \{ (Z_{Fi}, Z_{Ci}) : i = 1, \dots, M\}$ from (\ref{zDefs}) is a binomial likelihood $Z_{ki} \stackrel{\text{ind}}{\sim} \text{binomial}(n_{\text{ens}}, p_{ki})$ which represents a ``nonparametric'' approach to estimating the event probabilities, as no assumptions need to be made regarding the behavior of the underlying climate variable (as opposed to an extreme value distribution approach). A useful Bayesian framework for this likelihood involves a scenario-specific hierarchical model for the probabilities
\begin{equation} \label{hierP}
\begin{array}{c}
p_{ki} = \logit^{-1}( \mu_k + \beta_{ki})\\
\bfbeta_{k} \sim G_k
\end{array}
\end{equation}
for $k \in \{F, C\}$. Here, $\mu_k$ are scenario-specific (logit) means, $\bfbeta_k = (\beta_{k1}, \dots, \beta_{kM})$, and $G_k$ is a scenario-specific, mean-zero prior distribution for the region-specific effects that flexibly captures dependence among the regions. In principle, $G_C$ and $G_F$ need not be related; however, we model them as arising from the same class but allow for different hyperparameters (and hence the subscript). 

\begin{figure}[!t]
\begin{center}
\includegraphics[trim={0 0 0 0mm}, clip, width = \textwidth]{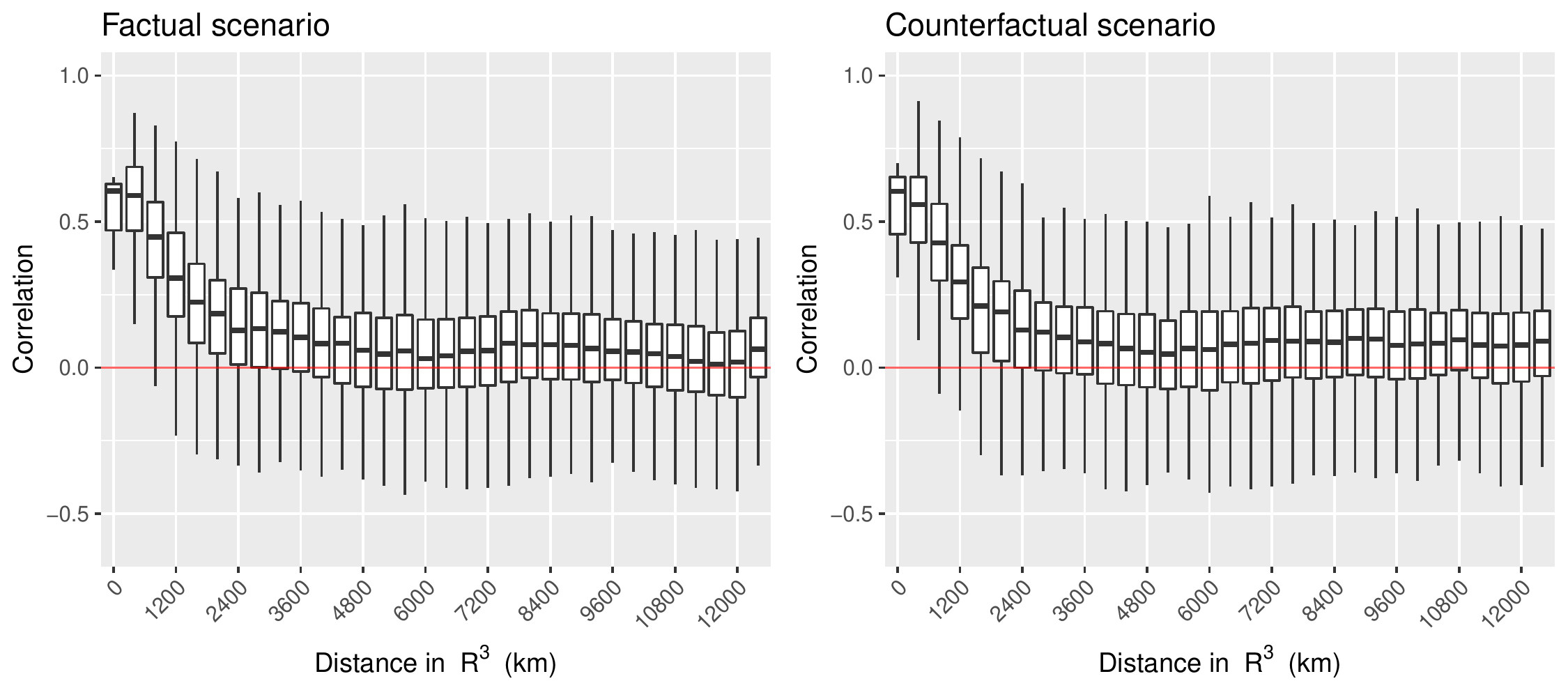}
\caption{Empirical correlation between the $\logit$ probability of a seasonally-adjusted hot January (1959-2014; on the anomaly scale) versus distance, for both the factual (left) and the counterfactual (right) scenarios.}
\label{correlation_hot_jan}
\end{center}
\end{figure}

While the literature contain a variety of options for how to model the $G_k$, existing approaches can neither flexibly model (potentially) non-Gaussian behavior nor directly account for irregular or long-range dependence due to teleconnections. As an illustration of the type of correlation we might expect to see in the scenario-specific probabilities, consider Figure \ref{correlation_hot_jan}, which shows box plots of the empirical correlations in the logit probability of a hot January over 1959-2014 (binned by distance; see below for details on how this is calculated). Note that the correlations for each scenario tend to be positive, even at long distances. Standard stationary spatial models will likely not be able to account for these irregular dependence relationships.

Therefore, we seek a model that more robustly uses available data to estimate the covariance between the hypotheses. One way to use prior knowledge to estimate the covariances between regions is to use long time series of climate model simulations: while the WRAF is generated based on monthly simulations of CAM5.1 (again see Section \ref{section42} for more details), there are also historical simulations of CAM5.1 available for both climate scenarios dating back to 1959. As such, we can use the empirical relationships between the historical simulations of both the factual and the counterfactual to inform the dependence relationships among the hypotheses. Formally, we can estimate monthly probabilities $\{ \widehat{p}_{ki}^{(t,j)}: k \in\{F, C\}; i = 1, \dots, M; t = 1, \dots, T; j = 1, \dots, 12\}$ ($t$ represents the year, $j$ represents the month) using a simple beta-binomial Bayesian model (maximum likelihood estimates are not used because zeros are possible), where the corresponding random variables $\{z_{ki}^{(t)}\}$ are calculated using a threshold specific to each month (note: the $z_{ki}^{(t)}$ are different than the random variables introduced in (\ref{zDefs})). Both the threshold for what is considered ``extreme'' and the count variables are calculated based on anomaly data (i.e., the atmospheric variables for each year are mean zero). Then, for a forecast in month $j$, we have an $M\times T$ matrix of probabilities $\widehat{\bf p}_k^{(j)}$ that can be used to calculate an empirical covariance on the logit scale:
$
\widehat{\bf S}^{(j)}_k = \Cov\big[ \logit \widehat{\bf p}_k^{(j)} \big]
$,
where $\widehat{\bf p}_k^{(j)} = \{ \widehat{p}_{ki}^{(t,j)}: i = 1,\dots,M; t = 1, \dots, T\}$. (Note: the correlation matrices used to create Figure \ref{correlation_hot_jan} are from the $\logit \widehat{\bf p}_k^{(1)}$ for hot months.)

Unfortunately, since for our application we have $T<M$ (the historical simulations only cover $T = 56$ years and there are $M=237$ regions), the resulting empirical estimate will not be a positive definite matrix; furthermore, it is well-known that the empirical covariance is a poor estimator for the true covariance (see, e.g., \citealp{Daniels2001}; \citealp{Bickel2008}). Instead, we can use a basis function approach where the basis functions are the eigenvectors of the estimated covariance $\widehat{\bf S}^{(j)}_k$, also known as empirical orthogonal functions (EOFs; see  \citealp{WikleHSS} or \citealp{CressieWikle}). EOFs are a popular strategy in modeling global climate variables, as the eigenvectors summarize the major modes of variability in a multivariate data set. Furthermore, it can be shown that the modes of variability (i.e., eigenvectors) are the same for the true covariance and a noisy estimate of the covariance (e.g., \citealp{CressieWikle}). The main idea here is to base the current forecast on past data. While the ``past'' (here, 1959-2014) is not necessarily a stationary climate (especially for the factual scenario), it can be argued that the modes of variability should be approximately consistent. As an example of the spatial patterns that we are able to capture using the EOF approach, consider the leading empirical EOFs for the logit probability of a hot January, shown in Figures B.3 and B.4 of Appendix B (in the supplemental materials).

Suppressing the $j$ notation, suppose for each month we have a set of $p$ EOF basis functions ${\bf h}_{kl} = \big( h_{kl}({\bf s}_1), \dots,  h_{kl}({\bf s}_M) \big)^\top, l = 1, \dots, p,$ for each scenario, collected into an $M\times p$ matrix ${\bf H}_k = ({\bf h}_{k1}^\top, \dots, {\bf h}_{kp}^\top)^\top$ (note that the EOFs are calculated separately for each scenario and event type). Then, following \cite{WikleHSS}, we can specify the following model for $\bfbeta_k = (\beta_{k1}, \dots, \beta_{kM})$:
\begin{equation}\label{EOFmodel}
\bfbeta_k = {\bf H}_k \boldsymbol{\alpha}_k + \boldsymbol{\xi}_k,
\end{equation}
where $\boldsymbol{\alpha}_k = (\alpha_{k1}, \dots, \alpha_{kp}) \in \mathbbm{R}^p$ is a random vector of basis function coefficients and $\boldsymbol{\xi}_k$ is a residual vector that captures discrepancies from the EOF basis function structure. Because the basis functions are orthogonal, the elements of $\boldsymbol{\alpha}_k$ can be considered independent \textit{a priori}. 

Using (\ref{EOFmodel}), we must specify three components: a prior distribution for the residual vector $\boldsymbol{\xi}_k$, a prior distribution for the basis function coefficients $\boldsymbol{\alpha}_k$, and the number of EOFs to include in the model (i.e., $p$). 

\subsection{Accounting for non-Gaussian discrepancy from the EOF structure} \label{section31}

If the number of EOFs is large enough to account for both large-scale and small-scale spatial variability, we can model the residual vector as independent and identically distributed (``iid'') random variables. A standard approach in nearly all statistical modeling is to assume that error is mean-zero and Gaussian, i.e., $\xi_{ki} \stackrel{\text{iid}}{\sim} N(0, \tau_k^2)$. However, in modeling extreme probabilities, such an assumption might be tenuous, even on the logit scale. Furthermore, in basing the dependence structure of the current forecast on past data, there is a risk of misspecifying the large-scale structure in the probabilities (that is, the EOFs). Therefore, it behooves us to use a more flexible approach in accounting for discrepancies from the fixed EOF structure.

One approach to more flexibly model the residual vector is via the skew-$t$ family of distributions (\citealp{Fernandez1998}; \citealp{Azzalini2003}; \citealp{FS2010}), which is a generalization of the Gaussian distribution that allows for skewness and heavy (non-exponential) tails. For a mean-zero random effect, this family involves three parameters: a scale parameter, a skewness parameter, and the degrees of freedom, which controls the heaviness of the tails. %; we write $\xi_{ki} \stackrel{\text{iid}}{\sim} ST(0, \tau_k^2, \omega_k, \nu_k)$. 
Actually, we use what \cite{ArellanoValle2008} call the ``centered'' parameterization for the skew-$t$ distribution, where $\sigma_k > 0$ is the scale parameter, $\delta_k \in (-1,1)$ controls the skewness, and $\nu_k>0$ is the degrees of freedom (see Appendix C in the supplemental materials for details); we write $\xi_{ki} \stackrel{\text{iid}}{\sim} ST(0, \sigma_k, \delta_k, \nu_k)$. 
 In any case, as the degrees of freedom $\nu$ approaches zero, the skew-$t$ distribution becomes very heavy-tailed and allows quite large deviations from zero. Also, note that the standard Gaussian error approach is a special case of the skew-$t$: for $\delta = 0$ and in the limit as $\nu \rightarrow \infty$, the residuals $\xi_{ki}$ are iid Gaussian.

\subsection{Sparsity-imposing prior for the EOF coefficients}

The EOF framework in (\ref{EOFmodel}) is a form of principal component regression (PCR), where the principal components (PCs) of a multivariate data set are used as regressors or covariates. In general, selecting an appropriate subset of PCs for PCR is extremely important for the sake of interpretation and parsimony of the resulting model; furthermore, we want to avoid overfitting the signal with too many PCs. Existing approaches for selecting the number of PCs (see \citealp{Jolliffe2002}) generally fall into one of three categories: graphical methods like the scree plot (\citealp{Cattell1966}); computational methods such as cross-validation (\citealp{Wold1978}; \citealp{Josse2012}); and model-based criteria like reversible jump MCMC (\citealp{Zhang2004}), marginal likelihood estimation (\citealp{Minka2000}), and model averaging (\citealp{Katzfuss2017}). Each of these approaches only consider ``nested'' PC models, in that a particular PC is included only if all lower-order PCs are included. 
% \cite{Wang2013} note that it is not always best to remove PCs with small eigenvalues, and furthermore the components with the largest eigenvalues may not be the most important in the PCR. 
Other approaches consider data-driven component selection for Bayesian PCR (\citealp{Wang2013}; \citealp{Lee2013}; \citealp{Junttila2015}), which use various prior distributions to regularize the PC coefficients. For example, \cite{Junttila2015} specify $\alpha_{kl} \stackrel{\text{iid}}{\sim} N(0, v)$ where the prior variance $v$ is estimated from the data, where the Gaussian prior corresponds to an $L_2$ penalty in penalized regression. Other recent papers by \cite{Hughes2013} and \cite{Guan2016} combine a Gaussian prior for the PC coefficients with either model selection or cross validation for choosing an appropriate number of PCs.

However, when dealing with empirical PCs, it is often the case that several of the PCs have extremely large variance (corresponding to large coefficients) and many have small variance (corresponding to small coefficients). In a penalized framework, an $L_2$ penalty corresponds to a linear smoother (see \citealp{Tansey2017}), which over-penalizes large signals and does not induce sparsity. Laplace priors, which correspond to an $L_1$ penalty, do encourage sparsity but still overshrink large signals in the presence of many near-zero signals due to their light tails. This is problematic in the PCR setting. A recent thread of research that addresses this problem in a Bayesian framework is the generalized double Pareto prior (GDP; \citealp{Armagan2013}), which has a spike at zero (like the Laplace prior) but heavy student's $t$-like tails. The GDP prior has a simple analytic form, yields a proper posterior distribution, and has a simple characterization as a scale mixture of Gaussian distributions: if $X \sim N(0, V)$, $V \sim \text{Exp}(U^2/2)$, and $U \sim \text{Gamma}(s, r)$, then the marginal distribution of $X$ is the GDP
\begin{equation} \label{GDP}
p(x | s, r) = \frac{r}{2s} \left(1 + \frac{|x|}{r}\right)^{-(s + 1)}, %\hskip4ex s , r > 0,
\end{equation}
written $X \sim GDP(s,r)$, where $s$ and $r$ are the Gamma shape and rate, respectively (\citealp{Armagan2013}). The GDP prior avoids the overshrinkage problems associated with Gaussian or Laplace priors (\citealp{Tansey2017}) and encourages sparsity (\citealp{Taddy2013}). (Note: related work by \cite{Polson2012} and \cite{Carvalho2010} also introduce heavy-tailed shrinkage priors with similar properties, but unfortunately these priors are not available in closed form when marginalized.)

Thus, while shrinkage priors have been used for PC or EOF selection, in a novel approach we propose to use the GDP prior as a more appropriate framework for incorporating EOF selection into the prior specification. Instead of worrying about how many EOFs to include in (\ref{EOFmodel}), we will instead incorporate all $T=56$ EOFs, so that $p=T$. Formally, the prior for the EOF coefficients is $\alpha_{kl} \stackrel{\text{iid}}{\sim} GDP(s,r)$ for $l = 1, \dots, T$. Note that using an exchangeable prior on the coefficients in this way ignores information about smoothness of the signals, an aspect which is not present in all regression settings but is present in PCR. When dealing with PCs or EOFs, we expect the variances of the empirical PCs to decay smoothly (following the eigenvalues), so that an exchangeable prior on the coefficients is not quite right. \cite{Lee2013} explicitly include a ``smooth'' prior for the coefficients by specifying a functional form for how their prior variances decay. However, in a general setting, it is not immediately obvious what type of decay is most appropriate. In our application, we use a different set of data to calculate the EOFs (i.e., the historical simulations) than what is used to actually estimate the coefficients (i.e., the year of simulations corresponding to the forecast year). In this case, there may be some mismatch between the historical simulations and the new data with respect to the magnitude and ordering of the signals.
% Furthermore, probabilistically speaking, the magnitude of the signals for any individual year might have a different ordering. 
Using an exchangeable prior can account for this mismatch and also does not require one to specify a functional form for the decay in the coefficient variances.

\subsection{Hyperprior specification and computation}

The hyperparameters for (\ref{EOFmodel}) are the mean $\mu_k$, the skew-$t$ parameters $\{ \sigma_k, \delta_k, \nu_k \}$ and the GDP parameters $\{s, r\}$. For the mean and skew-$t$ parameters we use proper but non-informative priors, namely
$p(\mu_k) = N(0, 10^2)$, $p(\sigma_k) = U(0, 100)$, $p(\delta_k) = U(-1, 1)$, and $p(1/\nu_k) = U(0,1)$,
where $U(a,b)$ is the uniform distribution over the interval $(a,b)$. Note that the prior for the degrees of freedom is actually on $1/\nu_k$ to improve mixing; furthermore, the upper bound of 1 on $1/\nu_k$ limits the tails to be no heavier than those of a Cauchy distribution. The hyperparameters of the GDP are slightly more complicated. \cite{Armagan2013} suggest fixing these at $s=1$ and $r=1$; \cite{Taddy2013} fix both hyperparameters at $s =1$ and $r = 1/2$, but found that his results were robust to other values of $s$. \cite{Tansey2017} also fix $s=1$ but encountered major problems when trying to estimate $r$ and instead fit separate models across a discrete grid of fixed values for $r$ and used DIC to choose the best value. Following these suggestions, we fix $s=1$, but, given the relative simplicity of (\ref{EOFmodel}) (compared to \citealp{Taddy2013} and \citealp{Tansey2017}), we were able to estimate $r$ from the data and used $p(r) = U(0, 100)$.

As is usually the case, the posterior distribution for this model is not available in closed form regardless of prior specification, so we resort to Markov chain Monte Carlo (MCMC) methods to obtain samples from the joint posterior distribution. We fit the model using the {\tt nimble} software for {\tt R} (\citealp{nimble}), which is a BUGS-like system for building and sharing analysis methods for statistical models, particularly for hierarchical frameworks. The MCMC is relatively straightforward, and code to fit the model is available in the online reproducibility documents.

%============================================
% SECTION 4: Model misspecification
%============================================
\section{Sensitivity to model misspecification} \label{section4}

The classical FDR procedures in the vein of \cite{SunCai2007} were developed for specific data models, and unfortunately \cite{Sun2015} find that the optimality of the procedure is quite sensitive to model misspecification. While the Bayesian procedures of \cite{Muller2004} are appropriate for more general classes of models, \cite{Newton2004} note that the bounds on $\overline{\FN}$, $\overline{\FD}$, and $\overline{\FDR}$ are ``approximate...because [they] rest on the accuracy of the fitted model.'' Furthermore, as noted in Section \ref{section2}, the performance of the Bayesian decision rules for Frequentist FDR is not guaranteed in the presence of correlation (\citealp{Pacifico2004}; \citealp{Guindani2009}).  As such, we wish to understand how both model misspecification and dependence impact the performance of these decision rules for the WRAF application and, subsequently, ensure that the nonparametric Bayesian framework developed in Section \ref{section3} is robust to both correlation and error in specifying a statistical model for the probabilities $\{ p_{ki} \}$, $ k\in\{F, C\}$. 

\begin{table}[!t]
\caption{The true states from which the simulated data sets are generated. Note: the Mat\'ern correlation function for GP-S and GP-L has smoothness $\nu = 2$. Furthermore, the EOF true states use a fixed number of EOFs ($p=30$) in the data generation.}
\begin{center}
\begin{tabular}{|c|c|}
\hline
\textbf{Label} 	& \textbf{True state} 									\\ \hline 
G-RE		& Gaussian random effects			 				\\ \hline
NG-RE		& Gamma (non-Gaussian) random effects					\\ \hline
GP-S		& (Mat\'ern) Gaussian process, short range of dependence	\\ \hline
GP-L			& (Mat\'ern) Gaussian process, long range of dependence		\\ \hline
EOF-G		& Fixed EOF structure, Gaussian discrepancy				\\ \hline
EOF-NG		& Fixed EOF structure, gamma (non-Gaussian) discrepancy	\\ \hline
\end{tabular}
\end{center}
\label{trueStates}
\end{table}%

To compare the performance of our robust nonparametric Bayesian approach outlined in Section \ref{section3} (henceforth labelled RNB) as well as several other related models within the various Bayesian decision rules based on $R_1$, $R_2$, and $R_3$, we perform a simulation study to explore the FDR performance for a variety of ``true'' states. For our simulation study, the number of regions will match that of the WRAF regions, i.e., $M=237$, and we use ensemble sizes of $n_{\text{ens}}=\{50, 100, 400\}$. A total of $N_{\text{rep}}=100$ data sets will be generated from each of six ``true states'' (see Table \ref{trueStates}) that are designed to represent the full space of all possible ``states'' for the CAM5.1 simulations. The hyperparameters for the true states will be fixed (see Appendix F of the supplemental materials), and the replicates will be drawn from the random effect distributions (as opposed to repeated binomial draws with the same effects). Complete details on the procedure for obtaining samples from each true state is provided in Appendix F, and code to generate samples from each true state is provided in the online reproducibility documents.	

For comparison, we also use a variety of standard models that might traditionally be used for the WRAF. Each of these again use the $Z_{ki}$ introduced in (\ref{zDefs}), as well as the binomial likelihood $Z_{ki} \stackrel{\text{ind}}{\sim} \text{binomial}(n_{\text{ens}}, p_{ki})$ used in Section \ref{section3}.

\vskip2ex
\noindent \textit{Classical likelihood ratio test}

\noindent In order to compare new approaches with classical FDR, we first outline a method for calculating a $P$-value for each null hypothesis using a Frequentist likelihood ratio test. Rewriting the hypotheses in terms of the probabilities $H: {p_{F}}/{p_{C}} \leq c$ (for now ignoring the region-specific subscript), the test statistic for a likelihood ratio test considers the ratio of likelihoods for $Z_C = z_C$ and $Z_F = z_F$:
\[
\lambda(z_C, z_F) = \frac{\sup_{\Theta_0} L(p_{F}, p_{C} | z_C, z_F)}{\sup_{\Theta} L(p_{F}, p_{C} | z_C, z_F)},
\]
where $\Theta_0$ is the parameter space defined by the null hypothesis and $\Theta$ is the entire (unrestricted) parameter space for $p_F$ and $p_C$. The likelihood is the product of individual Binomial likelihoods:

\[
L(p_F, p_C | z_C, z_F) \propto  \big( p_{F} \big)^{z_C} \big( 1 - p_F \big)^{n_C - z_C}  \big( p_C \big)^{z_F} \big( 1 - p_C \big)^{n_F - z_F}.
\]
It can be shown that the likelihood in the denominator is maximized for the MLEs $\widehat{p}_C = z_C/n_C$ and $\widehat{p}_F = z_F/n_F$. Alternatively, for the numerator, the restricted MLEs are
\[
\left(\widehat{p}^R_C, \widehat{p}^R_F\right) = \left\{ \begin{array}{lcl} (\widehat{p}_C, \widehat{p}_F) & \text{ if } & \widehat{p}_F > (\widehat{p}_C/c) \\
(\widetilde{p}_C, \widetilde{p}_C/c) & \text{ if } & \widehat{p}_F \leq (\widehat{p}_C/c),
\end{array} \right.
\]
where $\widetilde{p}_C = (1/4) \left( -b -\sqrt{b^2 - 8d} \right)$: $b = -[c (1 + \widehat{p}_F) + 1 + \widehat{p}_C]$, and $d=c (\widehat{p}_C + \widehat{p}_F)$ (\citealp{FarrMann1990}). Statistical theory says $-2\log \lambda(z_C, z_F) \stackrel{d}{\rightarrow} \chi^2_1$ as $n_C, n_F \rightarrow \infty$ ($\Theta$ involves two free parameters while $\Theta_0$ has just one); thus, an asymptotic $P$-value is $P\big( \chi^2_1 > -2\log \lambda(z_C, z_F)\big)$. Note that when $\widehat{p}_F > \widehat{p}_C/c$, the likelihood ratio is 1, $-2\log \lambda(z_C, z_F) = 0$, and the null hypothesis will never be rejected.
The resulting collection of $P$-values  can be used for a classical FDR (\citealp{BenjaminiHochberg1995}) or a Bonferroni-style family-wise error rate (FWER) procedure. %, which controls the probability of at least one false discovery.

\vskip2ex
\noindent \textit{Parametric Bayesian models for the risk ratio}

\noindent Again using the independent binomial likelihood for the $Z_{ki}$ in (\ref{zDefs}), the simplest Bayesian approach to modeling these probabilities is to estimate each of the $p_{Fi}$ and $p_{Ci}$ independently of each other and all of the other regions (an ``independent across regions'' model), henceforth M1. For $k\in\{F, C\}$ and $i=1, \dots, M$, simply use a conjugate beta prior $\pi(p_{ki}) = \mathcal{B}(a_p, b_p)$ for the binomial likelihood so that the posterior is 
\begin{equation} \label{indep_post}
\pi(p_{ki} | Z_{ki} = z_{ki}) = \mathcal{B}(z_{ki} + a_p, n_{\text{ens}} - z_{ki} + b_p).
\end{equation}
Posterior samples can be obtained by direct sampling from (\ref{indep_post}).  Alternatively, mirroring the hierarchical Bayesian framework in (\ref{hierP}), a variety of prior models for $G_k$ can be implemented:

\begin{itemize}
\item[M2] Exchangeable Gaussian prior: a random effects framework is useful for borrowing strength across the regions without the notion of spatial dependence. Here, we use $\beta_{ki} \stackrel{\text{iid}}{\sim} N(0, \tau^2_k)$. 

\item[M3] Exchangeable skew-$t$ prior: however, the Gaussian assumption may be too restrictive, in that the effects could be non-symmetric and heavy-tailed. Alternatively, we can use the skew-$t$ family of distributions (as in Section \ref{section31}), with $\beta_{ki} \stackrel{\text{iid}}{\sim} ST(0, \sigma_k, \delta_k, \nu_k)$. 

\item[M4] Conditionally autoregressive (CAR) prior: a natural approach for areal data like the WRAF regions, a CAR prior models $\boldsymbol{\beta}_k = (\beta_{k1},\dots, \beta_{kM})$ as a spatial random effect (see, e.g., \citealp{Banerjee2004} and \citealp{Pascutto2000}). Using a Gaussian model, the joint distribution for $\boldsymbol{\beta}_k$ can be defined in terms of the conditional distributions
\begin{equation*} \label{CAR}
p\big(\beta_{ki} | \{\beta_{kj}: j \neq i\} \big) = N \left( \frac{1}{|\partial i|} \sum_{j \in \partial i} \beta_{kj}, \frac{\tau_k^2}{|\partial i|} \right), \hskip3ex i = 1, \dots, M,
\end{equation*}
where $\partial i$ is the set of regions that share a border with region $i$ (the ``neighborhood'') and $|\partial i|= \#$ regions in the neighborhood. This specification is also called an \textit{intrinsic} CAR model which is an improper prior (\citealp{RueHeld2005} outline various ways to address this issue; see Appendix E).

\item[M5] Hybrid CAR/exchangeable prior: the model outlined in \cite{Leroux2000} offers a compromise between M2 and M4. As outlined in \cite{Leroux2000}, the $\tau^2_k$ parameter in M4 represents both overdispersion and spatial dependence, and these features may be in stark contrast. A variety of strategies are used in the literature to address this problem (see, e.g., \citealp{Cressie1991}; \citealp{Besag1991}); \cite{Leroux2000} instead specify an approach based on ``additive precisions,'' in which the precision matrix of the random effects is a convex combination of the exchangeable and CAR precision matrices:
\begin{equation} \label{Leroux_prec}
{\bf \Sigma}_k^{-1} \equiv \tau^{-2}_k \big[(1-\lambda_k) {\bf I} +  \lambda_k {\bf Q}\big]
\end{equation}
where ${\bf Q}$ is the CAR precision correlation matrix; $\lambda_k \in [0, 1]$ is a parameter that controls the degree of spatial dependence. Note that in this framework, $\lambda_k = 0$ corresponds to M2 while $\lambda_k = 1$ corresponds to M4; furthermore, ${\bf \Sigma}_k^{-1}$ is full rank for $\lambda \in [0,1)$. Using (\ref{Leroux_prec}), the full conditional distributions for the individual random effects are
\begin{equation*} \label{hybridCAR}
p\big(\beta_{ki} |  \{\beta_{kj}: j \neq i\} \big) = N \left(  \frac{\lambda_k}{1 - \lambda_k +\lambda_k|\partial i|} \sum_{j \in \partial i} \beta_{kj}, \frac{\tau_k^2}{1 - \lambda_k +\lambda_k|\partial i|} \right), \hskip3ex i = 1, \dots, M.
\end{equation*}

\item[M6] Gaussian process prior: another alternative to M4 is to use a Gaussian process prior for $\boldsymbol{\beta}_k$, defined for the centroids of each region (e.g., \citealp{Kelsall2002}). Like M5, independent random effects are a special (limiting) case of the Gaussian process prior, such that M6 can flexibly model both independent and dependent effects (unlike M4). For this approach, $\bfbeta_k \sim N_M({\bf 0}, {\bf \Sigma}_k)$: $\Sigma^{ij}_k = \tau_k^2 \mathcal{M}_\nu( ||\bfs_i - \bfs_j||/\phi_k)$,
where $\mathcal{M}_\nu(\cdot)$ is the Mat\'ern correlation function with smoothness $\nu$,  $\bfs_i, \bfs_j$ are the three-dimensional coordinates for the centroids of regions $i$ and $j$, and $||\cdot||$ represents Euclidean distance on $\mathbbm{R}^3$. In practice, since we are fitting a Gaussian process model to areal data and therefore do not observe data at very short distances, we fix $\nu = 0.5$.

\item[M7-9] EOF-based structure with a Gaussian prior for a fixed number of coefficients: for comparison, we use (\ref{EOFmodel}) with a more traditional exchangeable Gaussian prior on the coefficients, i.e., $\alpha_{kl} \stackrel{\text{iid}}{\sim} N(0, \sigma^2_\alpha)$, where $\sigma^2_\alpha$ is estimated from the data. In this framework, we use three different EOF truncations: $p=30$, which matches the number of EOFs used in the data generation (henceforth M7); $p=10$, where we use too few EOFs (henceforth M8); and $p=50$, where we use too many EOFs (henceforth M9). Models M7-9 will allow us to assess the performance of the GDP prior relative to more traditional priors with a specified truncation.

\end{itemize}
A summary of all the fitted models and their labels is given in Table \ref{modelFits}. Details on the hyperpriors and computation (via MCMC) are given in Appendices D and E (in the supplemental materials).

\begin{table}[!t]
\caption{A summary of the models fit to each simulated data set. Note that RNB and M1-M9 are implemented in a Bayesian framework.}
\begin{center}
\begin{tabular}{|c|l|}
\hline
\textbf{Label} 	& \textbf{Model description} 						\\ \hline \hline
RNB	& Robust nonparametric model with sparsity	\\ \hline\hline
LRT	& Classical likelihood ratio test (region-specific)			\\ \hline
M1	& Beta-binomial (independent-across-regions)			\\ \hline
M2	& Exchangeable Gaussian prior		 			\\ \hline
M3	& Exchangeable skew-$t$ prior						\\ \hline
M4	& Conditionally autoregressive (CAR) prior 			\\ \hline
M5	& Hybrid CAR/exchangeable prior					\\ \hline
M6	& Gaussian process prior with Mat\'ern correlation	\\ \hline
M7	& EOF-based structure with $p=30$ EOFs			\\ \hline
M8	& EOF-based structure with $p=10$ EOFs			\\ \hline
M9	& EOF-based structure with $p=50$ EOFs			\\ \hline
\end{tabular}
\end{center}
\label{modelFits}
\end{table}%

For each of the simulated data sets, the performance of the three decision criteria $R_1$, $R_2$, and $R_3$ will be compared by fitting each of the models outlined in Table \ref{modelFits}. The null hypothesis for each simulation will use a threshold value of $c = 1$ (testing for an increase in $p_F$ relative to $p_C$) while attempting to control FDR in ways comparable to the classical 0.1 significance level: for $R_1$, set $\alpha = 0.10$; for $R_2$, set $\lambda_2 = 1/0.1 -1 = 9$; for $R_3$, set $\gamma = 0.10M = 23.7$. Three different sets of hyperparameters will be used for each model, corresponding to cases in which most tests are true rejections (Scheme 1, $\approx0.85$), around half of tests are true rejections (Scheme 2, $\approx0.5$), and most tests are true nulls (Scheme 3, $\approx 0.15$). More details are provided in Appendix F.

Two final notes regarding the model fitting. First, for the Gaussian process model M6, note that the correlation function is set to be exponential, while the true states GP-S and GP-L have a Mat\'ern correlation with smoothness $\nu=2$. Second, all of the EOF approaches (RNB and M7-M9) require the initial step of estimating the EOF matrix, which is considered fixed when fitting the model. The EOFs used for true states EOF-G and EOF-NG are estimated from the historical simulations of temperature in January as described in Section \ref{section31} (the same EOFs are used for both generating data sets and model fitting). However, the benefit of the EOF framework is that it can robustly use available data to improve the model; therefore, when fitting RNB and M7-M9 to the other true states (G-RE, NG-RE, GP-S, and GP-L), we first calculate the EOFs using $T=56$ replicates drawn from the true state, separately for each scenario. For example, the EOFs used to fit data from GP-L would correspond to the covariance of a stationary Gaussian process with Mat\'ern correlation function. 

\vskip2ex
\noindent \textit{Results, summarized across simulated replicates}

%\clearpage
\begin{figure}[!t]
\begin{center}
\includegraphics[width = 0.94\textwidth]{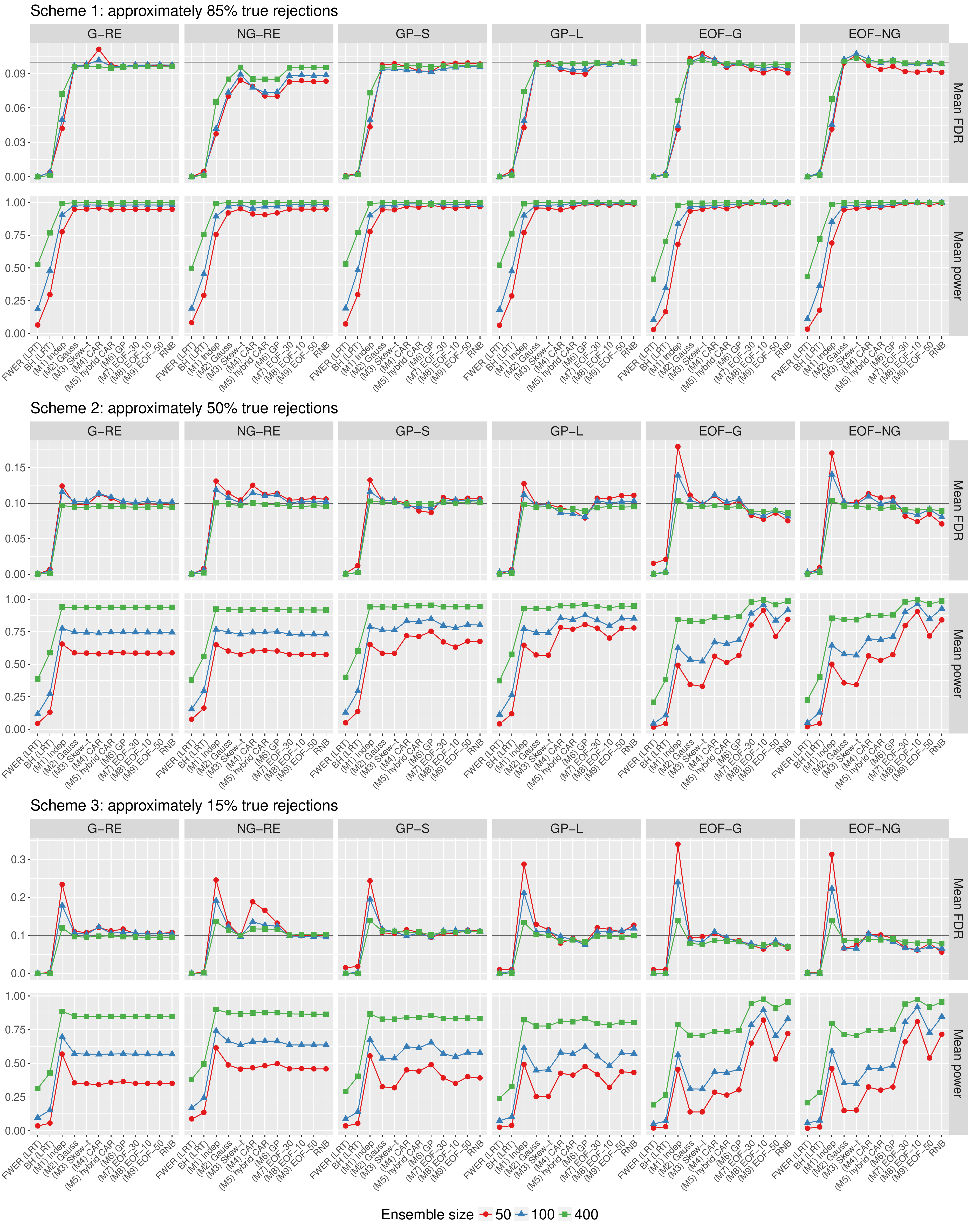}
\caption{FDR and power using the $R_1$ criteria, aggregated over the $N_{\text{rep}}=100$ replicates, for schemes 1, 2, and 3. Note that the $x$-axis in each subgrid corresponds to the different methods/fitted models. The target of $\alpha=0.1$ is plotted for FDR.}
\label{L1_criteria}
\end{center}
\end{figure}
%\clearpage

\noindent We present results for the $R_1$ criteria here, in the main text of the paper, as this decision criteria corresponds most closely with the classical notions of FDR; see Figure \ref{L1_criteria}. The top, middle, and bottom sub-plots show the FDR and power (i.e., the probability of rejecting a false null) for Schemes 1, 2, and 3 (respectively), averaged over the $N_{\text{rep}}=100$ replicated data sets. The sub-panels show the six true states, and the different methods/fitted models are shown along the $x$-axes. 

The first observation to make is that the Frequentist $P$-value approaches (FWER and BH) are clearly over-conservative, such that the realized FDR is approximately zero (well below the target $\alpha = 0.1$) across all schemes, ensemble sizes, and true states. This over-conservativeness shows up in the power plots as well, with the FWER procedure in particular suffering from extremely small power, even for the largest ensemble size. Interestingly, the Bayesian models RNB and M1-M9 each seem to do fairly well at controlling the FDR and maximizing the power (minimizing the FNR) for Scheme 1, across true states (aside from the independence model M1, which has somewhat reduced power particularly for the true states with spatial dependence). 

Schemes 2 and 3 tell a different story: for each of these schemes, and across true states, the independence model M1 is anti-conservative and fails to control the FDR (except for the largest ensemble size in Scheme 2). Otherwise, several items are noteworthy: the CAR model M4 performs poorly for the true states that do not include spatial dependence (G-RE and NG-RE); only the models that can accommodate skewness (models M3, M7-M9, and RNB) control the FDR for the NG-RE data. Otherwise, each of the models are mostly able to control the FDR, although major differences show up in the power. While, for example, M2 (a model without spatial dependence) is able to control the FDR for the GP-L simulations in Scheme 2, the power is significantly smaller than for a model that does accommodate dependence, e.g., M4. 

The EOF models M7-M9 and RNB perform comparably for the independent random effects (G-RE and NG-RE) with respect to both FDR and power, but yield major differences for the true states with spatial dependence. These differences are again most obvious in the power, where we can clearly see how under or overfitting the EOFs plays out. For the GP-S and GP-L effects, M8 (which uses only 10 EOFs) has reduced power relative to M7, M9, and RNB. This is not entirely surprising since 10 EOFs might be insufficient for characterizing a Mat\'ern covariance. For the EOF-G and EOF-NG true states, recall that these data were generated with 30 EOFs, so that M7 (which also uses 30 EOFs) is in a sense the ``correct'' model. However, as is discussed in Appendix F of the supplemental materials, to mimic the decay present in the corresponding empirical eigenvalues, the last 20 EOF coefficients have very small prior variance (see Table F.5 in the supplement). Therefore, M8 (with 10 EOFs) is also approximately the ``correct'' model. In fact, M8 performs best for the EOF true states, while overfitting the EOFs as in M9 results in greatly reduced power. Our new approach RNB nearly matches the power of M8 for these true states without having to specify an EOF truncation. Interestingly, the presence of spatial dependence (or lack thereof) in the simulated data has a larger effect on the power than the FDR: when the effects do not include dependence (G-RE and NG-RE), the power is roughly the same for models M2-M9 and RNB. This is true even for the NG-RE effects, for which the Gaussian-based models M2, M4, M5, and M6 struggle to control the FDR.

Therefore, if one were to choose a ``best'' model for the $R_1$ decision criteria, the robust nonparametric Bayesian model with a sparsity-imposing prior for the EOF coefficients and skew-$t$ discrepancy (RNB) is the clear choice, as it performs well across schemes and true states. RNB is able to control the FDR at approximately the nominal level for every combination of true state/scheme, and yields (almost) the largest power with the exception of the EOF-G/NG true states in Schemes 2 and 3. The only model that performs better for these true states is M8, which is suboptimal for the GP-S/L true states. Thus, RNB performs nearly as well as the best of the other EOF approaches, without requiring the specification of an EOF truncation. In some ways, this is not surprising, since the magnitude of the EOF coefficients together with their GDP shrinkage prior can differentiate between cases both with and without spatial dependence and the flexibility of the skew-$t$ residuals can capture both symmetric and non-symmetric effects. Furthermore, this approach allows us to more robustly use the data at hand (in this case, the historical CAM simulations) to capture irregular (nonstationary) spatial dependence patterns.

The story is largely the same for the $R_2$ and $R_3$ criteria (see Figures B.5 and B.6 in Appendix B of the supplemental materials): the RNB model yields the smallest loss (almost always), and controls the number of FDs while minimizing the number of FNs. Therefore, we have good reason to select the RNB model combined with the decision rule of interest as the procedure that best controls the realized loss, FDR, and FD.

%============================================
% SECTION 5: WRAF application
%============================================
\section{Applying the multiple testing procedure to the WRAF} \label{section5}

Having identified the robust nonparametric Bayesian model as an approach that flexibly controls false discoveries for each of the  procedures outlined in Section \ref{section2}, we now turn to applying the procedure to a real data set of climate model simulations.

\subsection{Selection of decision criteria} \label{sec41}

For this application, we decided not to use the second decision criteria ($R_2$) because it is not clear for the WRAF what the relative loss for each type of error should be; in other words, there is no obvious way to equate the cost of a false discovery and a false negative. In deciding between $R_1$ and $R_3$, we were initially drawn to $R_3$ because certain choices for the threshold $\gamma$ allow us to make sure our statements are scientifically significant. In systematically conducting a set of hypotheses regarding the presence of anthropogenic influence on extreme weather, in order to conclude a significant overall (global) influence we would need to reject a null hypothesis of no anthropogenic influence for some non-zero proportion of the globe; for example, we might want to see rejections for 5\% of the globe. In other words, from a practical perspective, we might be willing to make 10 false rejections (about 5\% of the 237 regions) because if we find fewer than 10 rejections then there is likely not a scientifically meaningful anthropogenic effect for the entire globe. Furthermore, in making an absolute (instead of a relative) statement about the number of false discoveries, the total number of discoveries relates to overall confidence: rejecting only a few hypotheses indicates low confidence that there is any overall anthropogenic effect, while rejecting many hypotheses indicates high confidence that there is indeed some overall anthropogenic effect.

However, upon further investigation, it became obvious that the $R_3$ criteria is too liberal. Returning to (\ref{BayesL3}), note that this procedure will always reject \textit{at least} $\left\lfloor \gamma \right\rfloor$ tests: in the most extreme case, where $\pi_{(i)} = 1$ for all $i$ (meaning the null hypothesis receives all of the posterior probability), we will still have
$
\sum_{i = 1}^{\left\lfloor \gamma \right\rfloor} \pi_{(i)} \leq \gamma$,
so that in this case $r_3 = \left\lfloor \gamma \right\rfloor$. In other words, a set of tests will be rejected, even though the posterior probability that each null hypothesis is true is 1; clearly, it is quite awkward to always reject a set of hypotheses despite the evidence. Therefore, if we use the $R_3$ criteria, after flagging a set of null hypotheses to reject we must then determine if the results are believable. For example, if $\gamma = 10$ and we only reject ten hypotheses, then we must conclude that almost all of these are false discoveries; alternatively, if we reject 100 hypotheses, then we can be confident that most of these are true rejections. However, what if we reject 15 hypotheses? Or 20? In these ``in-between'' cases, we must decide when enough tests have been rejected to conclude that at least some of the rejections are true.

The $R_1$ criteria, on the other hand, falls more in line with traditional multiple testing procedures, in that the conclusions drawn for a set of hypotheses are more appropriately adjusted for the fact that multiple tests are being conducted. Regardless of how many tests are rejected under this criteria, we can always be sure that (in expectation) only a small proportion of these are being falsely rejected. Furthermore, while the $R_3$ criteria might flag some hypotheses for rejection in spite of the large posterior probability that the null is true (see the previous paragraph), the $R_1$ criteria will only begin flagging hypotheses for rejection if the smallest posterior probabilities of the null being true (i.e., $\pi_{(1)}$, $\pi_{(2)}$, etc.) are close to zero. A final benefit of using the $R_1$ criteria is that the conclusions for nested hypotheses (see Section \ref{sec43}) will be consistent (e.g., the procedure will only reject $H_i: RR^{(\text{wet})}_i \leq 2$ if $H_i: RR^{(\text{wet})}_i \leq 1$ is also rejected), which is not the case for $R_3$. 

\subsection{Case study methods} \label{section42}

Having opted to use the $R_1$ criteria, we set $\alpha = 0.1$ (as is done in Section \ref{section4}). Practitioners often choose an FDR threshold based on common significance levels; here, we do the same, although there is no reason why this should be done (other than the fact that we want the FDR to be small but not too small such that we have no power). We then applied our robust nonparametric Bayesian model with the generalized double Pareto prior for the EOF coefficients with the $R_1$ criteria to the WRAF for two case studies: (1) hot events in January, 2015 and (2) wet events in March, 2015. For hot events, we use a more stringent cutoff for the null hypotheses, $c_{\text{hot}}=5$ (i.e., testing for $H_i: RR^{(\text{hot})}_i \leq 5$; this is due to the stronger anthropogenic signal for temperature), while for wet events we use $c_{\text{wet}}=1$. 

The data for estimating the $p_{ki}$ ($k \in \{F, C\}$) consist of output from large ensembles of simulations of version 5.1 of the Community Atmospheric Model (CAM5.1) global atmosphere/land climate model, run in its conventional $\sim 1^{\circ}$ longitude/latitude configuration (\citealp{NealeRB_ChenC-C_etalii_2012}; D. Stone et al., ``A basis set for exploration of sensitivity to prescribed ocean conditions for estimating human contributions to extreme weather in CAM5.1-1degree,'' submitted).  Simulations have been run under the experiment protocols of the C20C+ Detection and Attribution Project 
%\citep{StoneDA_PallP_2016}
(D. Stone and P. Pall, ``A benchmark estimate of the effect of anthropogenic emissions on the ocean surface,'' submitted), following two historical scenarios \citep{AngelilO_StoneD_etalii_2016} and will be regularly updated through time as a contribution to both the C20C+ D\&A project and the WRAF.  The first set of simulations (for the factual scenario) is driven by observed boundary conditions of atmospheric chemistry (greenhouse gases, tropospheric and stratospheric aerosols, ozone), solar luminosity, land use/cover, and the ocean surface (temperature and ice coverage).  The second set of simulations (for the counterfactual scenario) is driven by what observed boundary conditions might have been in the absence of historical anthropogenic emissions:  the anthropogenic component of atmospheric chemistry is set to year-1855 values, ocean temperatures are cooled by a seasonally- and spatially-varying estimate of the warming attributable to anthropogenic emissions, and sea ice concentrations are adjusted for consistency with the ocean temperatures (Stone and Pall, submitted).  Simulations within a scenario differ only in the starting conditions. The data and further details on the simulations are available at {\url{http://portal.nersc.gov/c20c}}. The simulations for both scenarios cover 01/1959 to 06/2015; the (time-varying) ensemble sizes are given in Table \ref{EnsSize}.
% is 50 for 01/1949--12/1996, 100 for 01/1997--12/2009, 400 for 01/2010--12/2013, 100 for 01/2014--10/2014, 98 for  10/2014--06/2015 in the factual (99 for 10/2014-06/2015 in the counterfactual). 

\begin{table}[!t]
\caption{CAM5.1 ensemble sizes from January, 1959 to June, 2015 used in the case studies.}
\begin{center}
\begin{tabular}{|c|c|}
\hline
Time range & Ensemble size \\ \hline \hline
01/1959 to 12/1996 & 50 (both factual and counterfactual) \\ \hline
01/1997 to 12/2009 & 100 (both factual and counterfactual) \\ \hline
01/2010 to 12/2013 & 400 (both factual and counterfactual) \\ \hline
01/2014 to 10/2014 & 100 (both factual and counterfactual) \\ \hline
10/2014 to 06/2015 & 98 (factual) \\ \hline
10/2014 to 06/2015 & 99 (counterfactual) \\ \hline
\end{tabular}
\end{center}
\label{EnsSize}
\end{table}%

The event of interest for both case studies (used to define the region-specific probabilities $p_{ki}$) is the occurrence of a month that is more extreme than the third most extreme event expected over the preceding 30 year period. In other words, for a forecast in 2015, the event definition is the 1-in-10 year event which, for hot and wet months, corresponds to the 0.9 quantile of average monthly temperature or precipitation for each region in the factual simulations over 1985-2014 (specifically using the monthly measurements from the 50-member ensemble that covers this entire period). Using a moving time period of fixed length (30 years) ensures that we have accounted for climate change and that the events we consider are extreme in the ``current'' climate. All of the historical CAM simulations (both the factual and counterfactual) from the entire 1959-2014 period are used to calculate the EOFs following the procedure outlined in Section \ref{section3}; the simulations from 2015 are used to fit the statistical model and classify the hypotheses. Otherwise, all prior specifications and computation via MCMC are the same as described in the simulation study. Note that an implicit assumption of this application is that the CAM5.1 simulations are suitable for evaluating changes in the probability of extremes \citep{AngelilO_PerkinsKirkpatrickS_etalii_2016,AngelilO_StoneD_etalii_2016}.

\subsubsection{Results for a single set of hypotheses}

\begin{figure}[!t]
\begin{center}
\includegraphics[width = \textwidth]{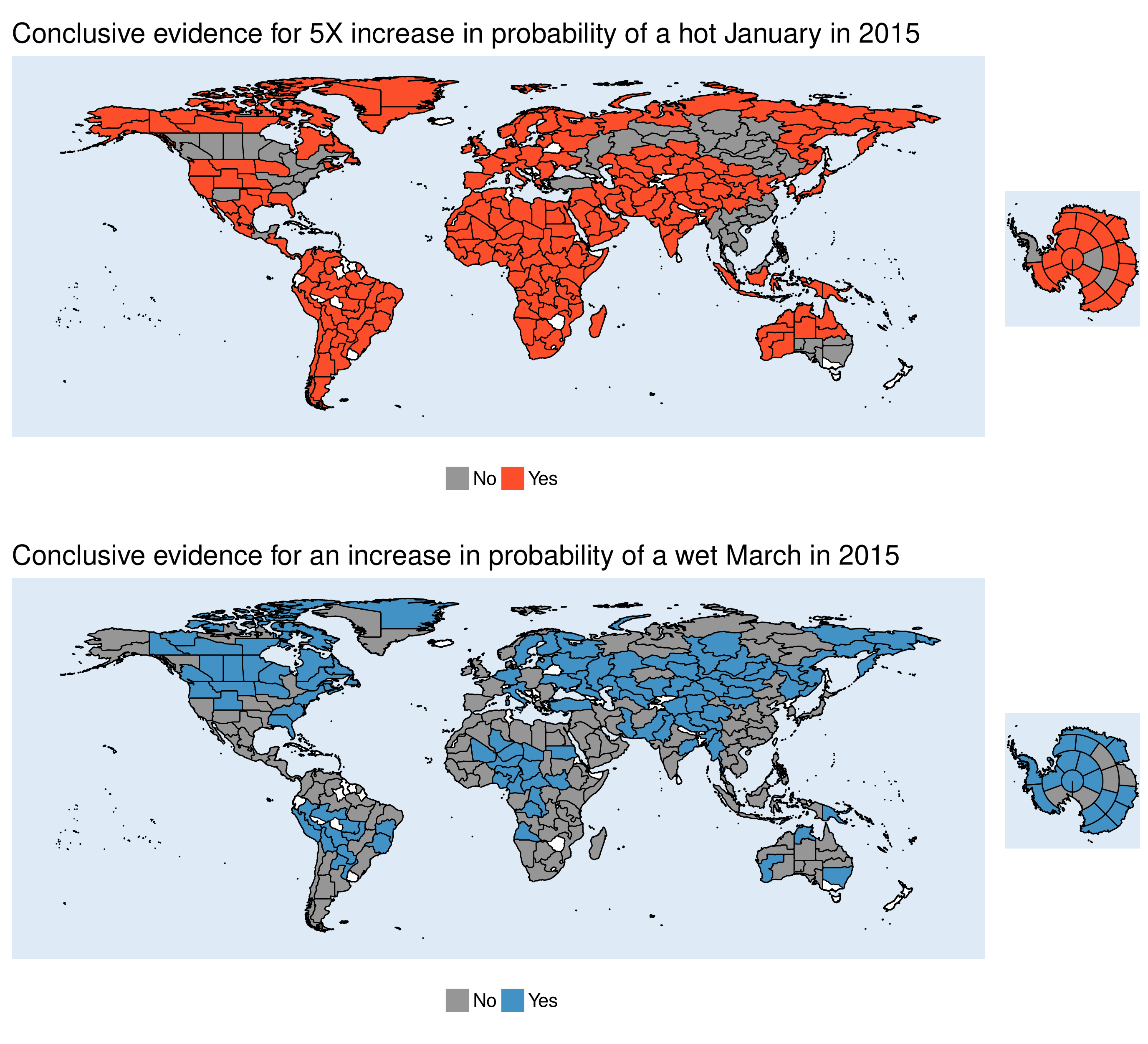}
\caption{Results of testing a collection of hypotheses $H_i: RR^{(\text{hot})}_i \leq 5$ (top; i.e., determining if there is conclusive evidence for a five-fold increase in the probability that January, 2015 will have an average temperature that exceeds the third hottest expected January over 1985-2014) and $H_i: RR^{(\text{wet})}_i \leq 1$ (bottom; i.e., determining if there is conclusive evidence for an increase in the probability that the total precipitation in March, 2015 will exceed the third wettest expected March over 1985-2014). The white areas (e.g. New Zealand, Zimbabwe) do not satisfy criteria for fitting into political regions of the target 400,000-900,000 km$^2$ range as described in Stone (submitted) and are not analyzed here.
}
\label{hotJan_wetMarch}
\end{center}
\end{figure}

The results for each case study are shown in Figure~\ref{hotJan_wetMarch}. Even with a larger cutoff for hot Januarys (testing for a five-fold increase as opposed to simply an increase), an overwhelming majority of the regions (194 of 237) have experienced a large degree of anthropogenic warming in 2015, with only a few regions in North America, Southeast Asia, central Russia, and southeast Australia failing to provide conclusive evidence of a five-fold increase in occurrence probability of a hot January, 2015 (every region has conclusive evidence against the null hypothesis when the cutoff is relaxed to $c_{\text{hot}} = 1$). The results are more varied for wet events in March of 2015, as there are many regions with and without conclusive evidence against the null hypothesis.  Many regions in the northern extratropics (mid to high latitudes) have an increased probability of a wet event in March, 2015 as a result of anthropogenic emissions.  An increased probability is the general tendency along an equatorial band as well (although not in Southeast Asia), while the subtropics (arid regions in the tropics) generally lack conclusive evidence;  a notable exception is over the northern Sahel, which may indicate an earlier advance of the West African monsoon in this climate model due to anthropogenic emissions \citep{LawalKA_AbatanAA_etalii_2016}.

\begin{figure}[!t]
\begin{center}
\includegraphics[width = \textwidth]{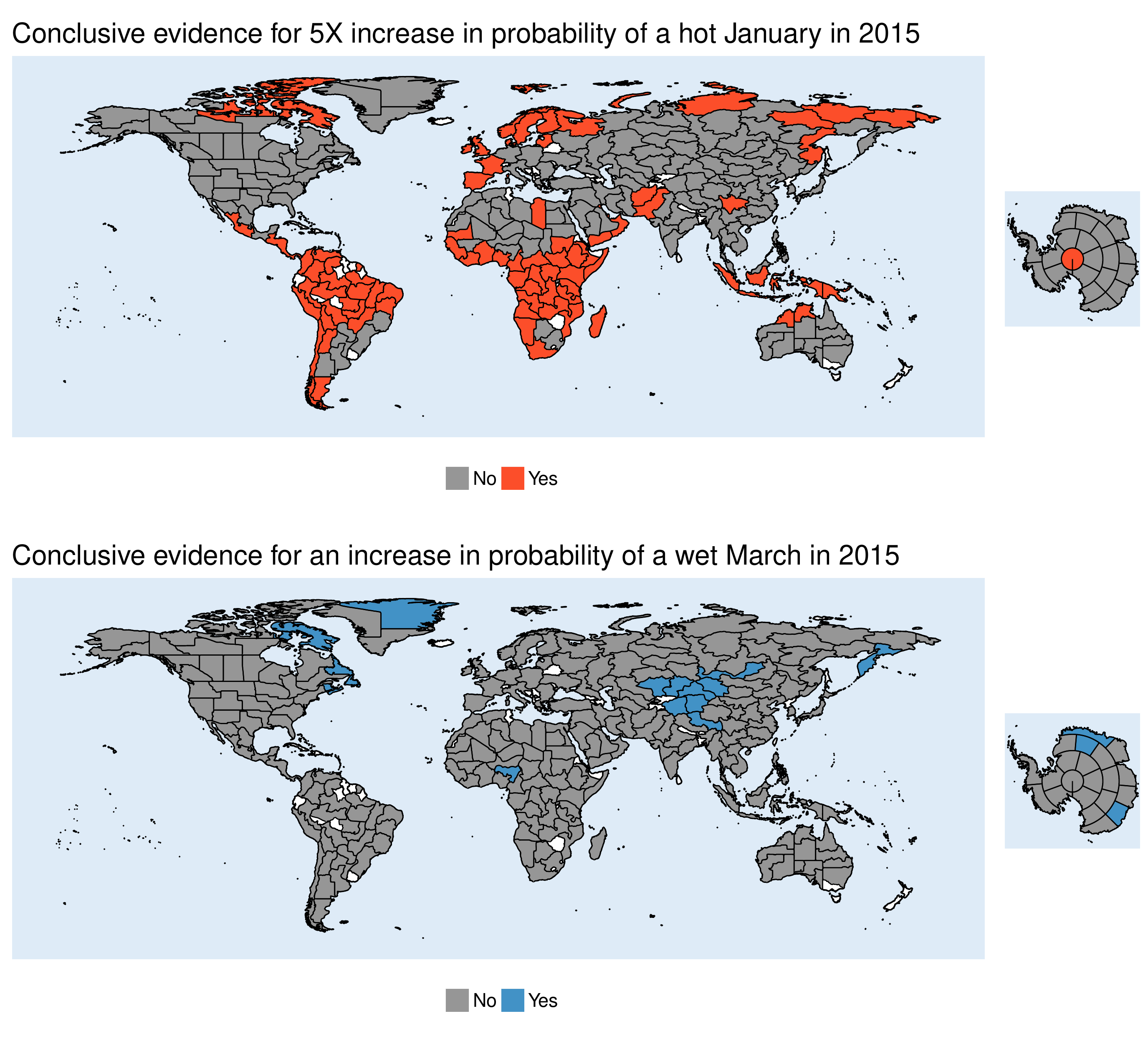}
\caption{As in Figure \ref{hotJan_wetMarch}, but using the \cite{BenjaminiHochberg1995} procedure based on likelihood ratio test $P$-values. The white areas (e.g. New Zealand, Zimbabwe) do not satisfy criteria for fitting into political regions of the target 400,000-900,000 km$^2$ range as described in Stone (submitted) and are not analyzed here.
}
\label{BH_hotJan_wetMarch}
\end{center}
\end{figure}

For comparison to standard (frequentist) FDR methods, we show maps for the corresponding forecast using the traditional \cite{BenjaminiHochberg1995} procedure based on likelihood ratio test $P$-values; see Figure \ref{BH_hotJan_wetMarch}. The BH procedure is again much more conservative than the corresponding results using our Bayesian approach (shown in Figure \ref{hotJan_wetMarch}), identifying conclusive evidence of changes in extreme probabilities for a greatly reduced subset of the WRAF regions. As an aside, we note that maps like Figure \ref{hotJan_wetMarch} produced using the Bayesian framework with other fitted models (i.e., M2-M9; not shown) yield only mild differences from the map based on our new modeling approach. In this case, where there is no way to assess which fitted model yields the ``correct'' results, we prefer our new approach based on the results of the simulation study in Section \ref{section4}.

\subsubsection{Capturing the existence and magnitude of anthropogenic influence} \label{sec43}

Both the current and planned upcoming versions of the WRAF actually conduct more than one set of hypothesis tests for each forecast: several different thresholds are used (e.g., $c_{\text{wet}} = 1$ versus $c_{\text{wet}} = 2$) in conjunction with several different types of null hypotheses (e.g., $H_i: RR^{(\text{wet})}_i \leq c_{\text{wet}}$ versus $H_i: RR^{(\text{wet})}_i \geq c_{\text{wet}}$). The purpose of these categories is to make statements that combine confidence in the change in probability as well as the magnitude of this change. As such, the forecast actually involves ``multiple-multiple testing,'' in that we now have multiple sets of $M$ hypotheses to test. This can be accomplished in our framework by simply conducting the classification procedure several times; recall from Section \ref{sec41} that using $R_1$ yields consistent results for nested hypotheses (unlike $R_3$). The testing adjustment is done separately for each category, and therefore the existence of any possible false discoveries can be interpreted within each category.

\begin{figure}[!t]
\includegraphics[width = \textwidth]{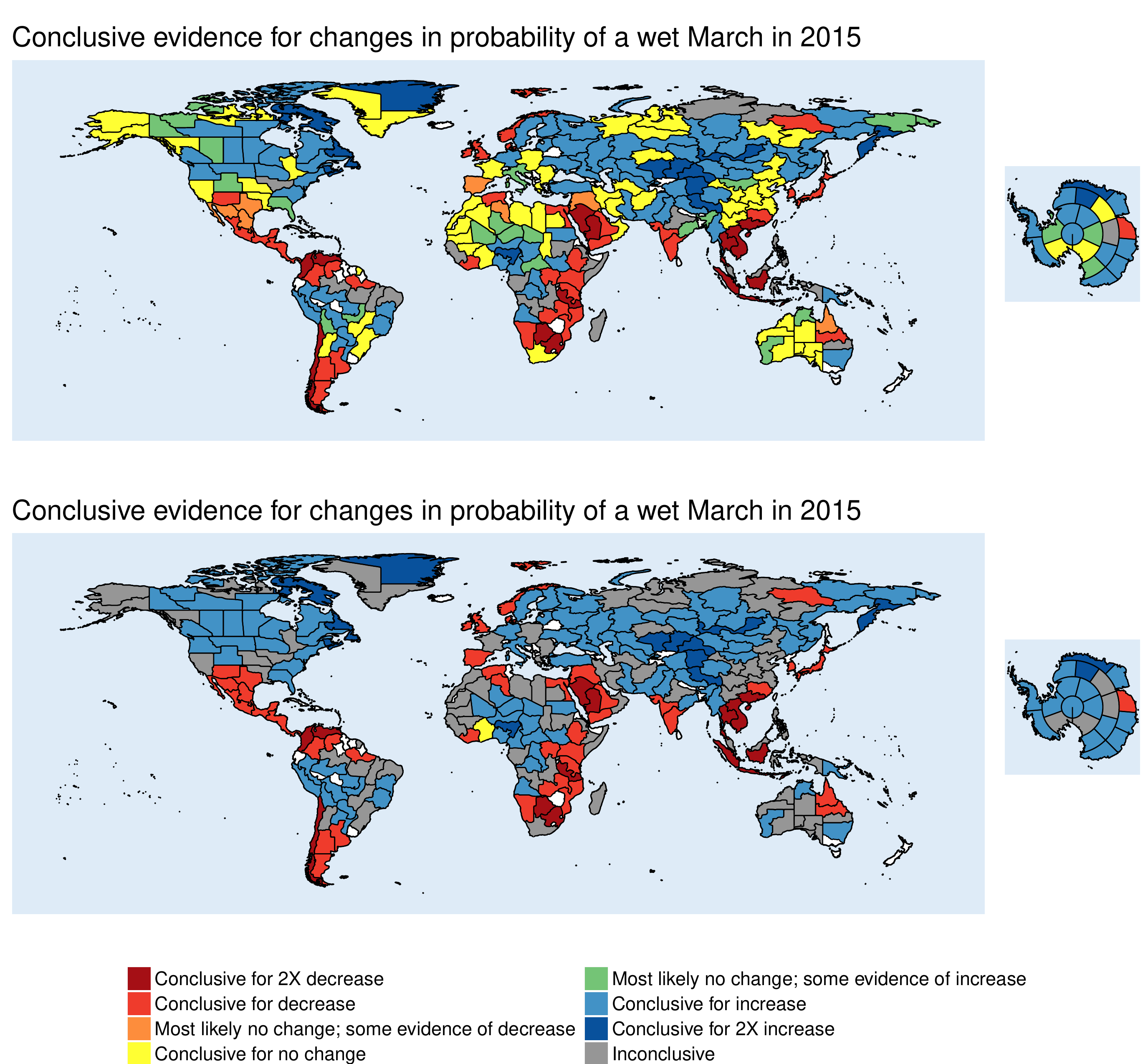}
\caption{Results of testing multiple hypotheses per region, in order to capture the magnitude and direction of the effect of anthropogenic influence. 
Top: $1/2 \leq RR_i \leq 2$ defines the ``Conclusive for no change'' category; bottom: $2/3 \leq RR_i \leq 3/2$ defines the ``Conclusive for no change'' category.}
\label{wetMarch_mult}
\end{figure}

As an example of what the attribution forecast looks like for multiple categories, see Figure \ref{wetMarch_mult}. A benefit of the Bayesian framework is that we can first test for the \textit{absence} of an anthropogenic effect using a null hypothesis like
$
H^{(1)}_i: RR^{(\text{wet})}_i \leq l_{\text{absence}} \cup RR^{(\text{wet})}_i \geq u_{\text{absence}}
$,
where $u_{\text{absence}}$ and $l_{\text{absence}}$ are upper and lower limits, respectively, for an interval including $1$ that defines ``no anthropogenic influence.'' Regions where we can reject $H^{(1)}_i$ display strong evidence that anthropogenic forcings have not changed the probability of extreme precipitation. Otherwise, the other null hypotheses of interest are 
\[
H^{(2)}_i: RR^{(\text{wet})}_i \geq 1/2; \hskip2ex H^{(3)}_i: RR^{(\text{wet})}_i \geq 1; \hskip2ex H^{(4)}_i: RR^{(\text{wet})}_i \leq 1; \hskip2ex H^{(5)}_i: RR^{(\text{wet})}_i \leq 2;
\]
being able to reject these hypotheses indicates conclusive evidence that the probability of extreme precipitation is decreased by a factor of two, decreased, increased, or increased by a factor of two (respectively). There is clearly some overlap between $H^{(1)}_i$ and both $H^{(3)}_i$ and $H^{(4)}_i$; to reflect this, we create two additional categories 
%labelled on Figure \ref{wetMarch_mult},
to indicate regions that reject both $H^{(1)}_i$ and $H^{(3)}_i$ (orange, indicating that while there is most likely no change in the probability there is some evidence for a decrease) as well as both $H^{(1)}_i$ and $H^{(4)}_i$ (green, indicating that while there is most likely no change in the probability there is some evidence for an increase). A final category (shown in gray) identifies regions that fail to reject any of the hypotheses and are thus classified as inconclusive. 

Maps of these multi-category results are shown in Figure \ref{wetMarch_mult}, where we use both a wide interval $l_{\text{absence}} = 1/2$ and $u_{\text{absence}} = 2$ as well as narrower limits $l_{\text{absence}} = 2/3$ and $u_{\text{absence}} = 3/2$. \cite{AngelilO_StoneD_etalii_2016} provide justification for using the narrower interval $(2/3, 3/2)$ as the definition of ``no anthropogenic influence''; however, the somewhat limited ensemble sizes ($\approx 100$) in this case study prevent us from conclusively finding no change (except in one region) for the narrower interval (bottom, Figure \ref{wetMarch_mult}). The wider interval used for the top panel of Figure \ref{wetMarch_mult}, on the other hand, concludes that a fairly large proportion of the map experiences no change. Clearly, being able to conclude that extreme probabilities are unchanged between the two climate scenarios depends heavily upon both the ensemble size and width of the interval that defines ``no change,'' and this tradeoff can result in significant qualitative differences. Given that the geometric range spanned by $(2/3, 3/2)$ is about half that spanned by $(1/2, 2)$, a quadrupling of the ensemble size (i.e., increased to 400 members) would be expected to result in the identification of a substantial number of regions in the ``no change" categories, in contrast to when only 100 members are available (as also noted in analysis of 2013 events when the ensemble sizes are larger; not shown).

%============================================
% SECTION 6: Conclusion/discussion
%============================================
\section{Conclusions} \label{section6}

In this paper, we have developed a hierarchical Bayesian modeling framework for estimating the probability of extreme events and the risk ratio over a large collection of land-regions, as well as a decision theoretic procedure that allows us to flexibly control the number of false discoveries while maximizing the number of true discoveries. The Bayesian hierarchical model robustly uses historical climate model simulations to estimate irregular (nonstationary) dependence patterns among the hypotheses, can account for non-Gaussian behavior in the region-specific effects, and uses an appropriate shrinkage prior that does not require choosing an EOF truncation point. Furthermore, we show that the modeling framework maintains false discovery control even when the true data-generating mechanism arises from a completely different class of statistical models. Finally, we apply our robust statistical model to a real data set used for making seasonal forecasts for the Weather Risk Attribution Forecast. Moving forward, we plan to operationalize our procedure as described in Section \ref{section42} to replace the current ad hoc presentation of the forecast.

We have demonstrated the application of our procedure across regions and with multiple hypotheses for each region.  However, we have not applied it across event types (e.g. hot, cold, wet, and dry events for a single region) or across multiple months.  Application across regions makes sense for several reasons.  First, events are presented in global maps of these regions (as in Figure \ref{wetMarch_mult}) and thus are not only providing information on each region individually but also on the aggregate of all of the regions.  Second, even with some correlation across the regions, there remains a large ``effective sample size'' of tests, whereas testing across event types would yield a small number of tests (such that a multiple testing adjustment has less value).  While testing across multiple months (e.g., all months or only the same calendar month from a given period of years) may provide a moderate number of tests, it would be hard to fit into the monthly operational design of the WRAF. Continual updating of past calculations, as further months become available, would pose a presentation and communication challenge. However, in a more retrospective research framework, studying events over a decade for instance, testing across months as well as, or instead of, regions could make sense.

Finally, while the final version of the forecast will use the $M=237$ regions shown in Figure \ref{hotJan_wetMarch} (where each region is approximately 0.5 million km$^2$; these are the WRAF05 regions), the WRAF will also provide a forecast for aggregates of these regions: 68 regions comprising 2 million km$^2$ each (WRAF2); 30 regions comprising 5 million km$^2$ each (WRAF5); and 12 regions comprising 10 million km$^2$ each (WRAF10). As a demonstration of how our procedure will perform for a smaller number of regions, we conducted a simulation study similar to the one in Section \ref{section4} using the larger WRAF2 regions ($M=68$; these regions are slightly modified from the current version of the WRAF, which has 58 regions); these results are shown in Appendix A.2 (see supplemental materials). Results for the WRAF2 regions are approximately consistent with the simulation study results for the smaller, WRAF05 regions.

\if1\blind{
%============================================
% References
%============================================
\section*{Acknowledgements}
This research was supported by the Director, Office of Science, Office of Biological and Environmental Research of the U.S. Department of Energy under Contract No. DE-AC02-05CH11231 % as part of their Regional \& Global Climate Modeling Program (RGCM) 
and used resources of the National Energy Research Scientific Computing Center (NERSC), also supported by the Office of Science of the U.S. Department of Energy, under Contract No. DE-AC02-05CH11231.

} \fi

\if0\blind{
} \fi

%============================================
% References
%============================================
\bibliographystyle{apalike} \bibliography{FDR_literature}

\clearpage
%============================================
% Appedices
%============================================
%============================================
% Appendix
%============================================
\begin{appendix}
\numberwithin{equation}{section}
\numberwithin{figure}{section}
\numberwithin{table}{section}

% Section 2 ====================================
\section{Classical model-specific decision theory approaches for false discovery control} \label{section21}

Using a Frequentist perspective, \cite{SunCai2007} frame the multiple testing problem in a compound decision theory framework. This thread of research considers controlling the marginal FDR, using the fact that, under weak conditions, $\mFDR = \mathsf{E}(\FDR) + \mathcal{O}(M^{-1/2})$ (\citealp{Genovese2002}). \cite{SunCai2007} note that two approaches can be taken to address the multiple testing problem. First, one can set out with the goal of separating the non-null hypotheses from the nulls, using a weighted classification approach. In other words, the decision rule $\boldsymbol{\delta}$ is constructed by minimizing the classification risk $\mathsf{E}[L_\lambda(\bftheta, \boldsymbol{\delta})]$, where the loss function is
\begin{equation} \label{FreqLoss}
L_\lambda(\bftheta, \boldsymbol{\delta}) = \frac{1}{M} \sum_{i=1}^M \Big\{ \lambda (1-\theta_i)\delta_i + \theta_i(1-\delta_i)   \Big\};
\end{equation}
here, $\lambda>0$ is the loss attached to a false positive error (relative to a false negative error). Alternatively, one can set out with the goal of discovering as many true findings as possible while incurring a low proportion of false positive findings: in other words, find $\boldsymbol{\delta}$ with the smallest false non-discovery rate (FNR) among all rules with the FDR bounded by $\alpha\in (0,1)$. \cite{SunCai2007} go on to show that these two approaches are equivalent as long as a monotone likelihood ratio condition is satisfied; that is, the optimal solution to the classification problem (where $\lambda$ depends on the desired $\alpha$) is also optimal for the multiple testing approach, in the sense that the classification rule yields the smallest marginal false negative rate ($\mFNR$) among all procedures that bound $\mFDR \leq \alpha$.

Unfortunately, proofs for the optimality of all of these procedures rely on the notion of independent hypotheses, and the optimality is called into question when the hypotheses are instead dependent. On one hand, \cite{BenjaminiYekutieli2001} show that FDR is controlled at the stated level for dependent hypotheses using either the original approach in \cite{BenjaminiHochberg1995} or the adaptive procedure in \cite{Benjamini2000}. However, on the other hand, \cite{Efron2007} found that non-zero correlation between tests can result in testing procedures that are either too conservative or too anti-conservative; \cite{Schwartzman2011} show that the procedure can fail to be consistent as the number of tests grows under certain types of dependence. \cite{SunCai2009} also note that in dealing with the effects of correlation on an FDR procedure, the efficiency of the procedure should be the focus (not just the validity), and that failing to model any known dependence structure can impact the optimality of the procedure. The decision rules of \cite{BenjaminiHochberg1995}, \cite{Benjamini2000}, \cite{Efron2001}, and \cite{SunCai2007} are simple, meaning that $\delta_i$ is a function only of $Z_i$; i.e., $\delta_i({\bf Z}) = \delta_i(Z_i)$, and therefore symmetric, meaning that $\boldsymbol{\delta}(\tau({\bf Z})) = \tau(\boldsymbol{\delta}({\bf Z}))$ for all permutation operators $\tau$ (\citealp{SunCai2007}). It is easy to imagine that in the case of correlated hypotheses, compound decision rules (i.e., decision rules $\boldsymbol{\delta}$ such that $\delta_i$ depends on the other $Z_j$, $j\neq i$) are preferred in that they might be able to identify non-nulls with a smaller signal by pooling information across tests. For example, when hypotheses are positively correlated within a temporal or spatial domain, one would expect that the non-null $\theta_i$ would appear in groups or clusters (\citealp{SunCai2009}). 

As a result, \cite{SunCai2009} extend the compound decision framework for multiple testing in the presence of dependence. Specifically, modeling the unknown $\theta_i$ as random effects arising from a hidden Markov model (HMM), \cite{SunCai2009} prove that the optimal classification rule for the loss function (\ref{FreqLoss}) is of the form $\delta_i = I({T}_i < t_\lambda)$, where 
\begin{equation} \label{oracleStat}
T_i = P_{{\boldsymbol{\xi}}}(\theta_i = 0 | {\bf Z})
\end{equation}
is the so-called ``oracle statistic'' and $\boldsymbol{\xi}$ is a vector of all hyperparameters in the HMM. It is important to note that the derivation of (\ref{oracleStat}) in \cite{SunCai2009} as the oracle statistic is specific to the HMM framework. Furthermore, because the HMM satisfies a monotone likelihood ratio condition, $T_i$ is also the optimal statistic for the multiple testing problem, in that $\delta_i = I({T}_i < t_\lambda)$ yields the smallest $\mFNR$ subject to $\mFDR \leq \alpha$. The relationship between $\lambda$ and $\alpha$ can be seen by writing the decision rule as a step-up procedure (like \citealp{BenjaminiHochberg1995}): first, rank the oracle statistics ${T}_{(1)} \leq \cdots \leq {T}_{(M)}$, and find
\begin{equation} \label{SunCai}
r = \max\left\{ j: \frac{1}{j} \sum_{i=1}^j {T}_{(i)} \leq \alpha\right\};
\end{equation}
then, reject $H_{(1)}, \dots, H_{(r)}$. In practice, of course, the $T_i$ (and hence the $\{\delta_i\}$ and $r$) are unknown: \cite{SunCai2009} outline a data-driven procedure that uses a plug-in estimate $\widehat{\boldsymbol{\xi}}$ to estimate $\widehat{T}_i = P_{\widehat{\boldsymbol{\xi}}}(\theta_i = 0 | {\bf Z})$ and therefore determine  $r$ by replacing $T_{(i)}$ with $\widehat{T}_{(i)}$ in (\ref{SunCai}). Since the estimated oracle test statistic for the $i$th hypothesis depends on the entire vector of data, \cite{SunCai2009} note that the decision rule is neither simple nor symmetric. 

Two recent papers by \cite{Sun2015} and \cite{Shu2015} extend the work of \cite{SunCai2009} to provide similar results for spatial random fields and multi-dimensional Markov random fields (MRFs), respectively. In spite of the different statistical models, in both cases the oracle statistic is the same as (\ref{oracleStat}) and the decision rule can be written as (\ref{SunCai}). However, model-specific proofs are required to verify that (1) the classification risk is indeed minimized by $\delta_i = I({T}_i < t_\lambda)$, and (2) the optimal classification (oracle) statistic satisfies a monotone likelihood ratio condition and hence yields the smallest $\mFNR$ among all procedures with $\mFDR\leq\alpha$ (here, both $\mFNR$ and $\mFDR$ are defined in a Frequentist sense). Furthermore, \textit{estimation} of the oracle statistic $T_i$ is, of course, model-specific. \cite{SunCai2009} use random effect prediction conditional on hyperparameter estimates: in the HMM, conditional on $\widehat{\boldsymbol{\xi}}$, the oracle statistic can be expressed in terms of forward and backward density variables, which can be calculated recursively.  \cite{Sun2015} also conduct random effect prediction (albeit marginalizing over hyperparameters), but, since there is no longer an iterative formula for calculating the $\widehat{T}_i$ for a Gaussian random field, they instead utilize the Bayesian computational framework (i.e., Markov chain Monte Carlo) as a way to ``extract information effectively from large spatial data sets'' and implement their data-driven procedure.

Both \cite{SunCai2009} and \cite{Sun2015} conduct simulation studies to verify that their approach outperforms traditional FDR procedures (e.g., BH and AP) when simulated data arise from the true statistical model (i.e., HMM or Gaussian random field). However, \cite{Sun2015} also find that ``the precision of [their] testing procedure shows some sensitivity to model misspecification.''

% Supplemental Figures ===========================
\section{Supplemental figures} \label{SuppFig}

\begin{figure}[!h]
\begin{center}
\includegraphics[trim={45 15 45 25mm}, clip, width = \textwidth]{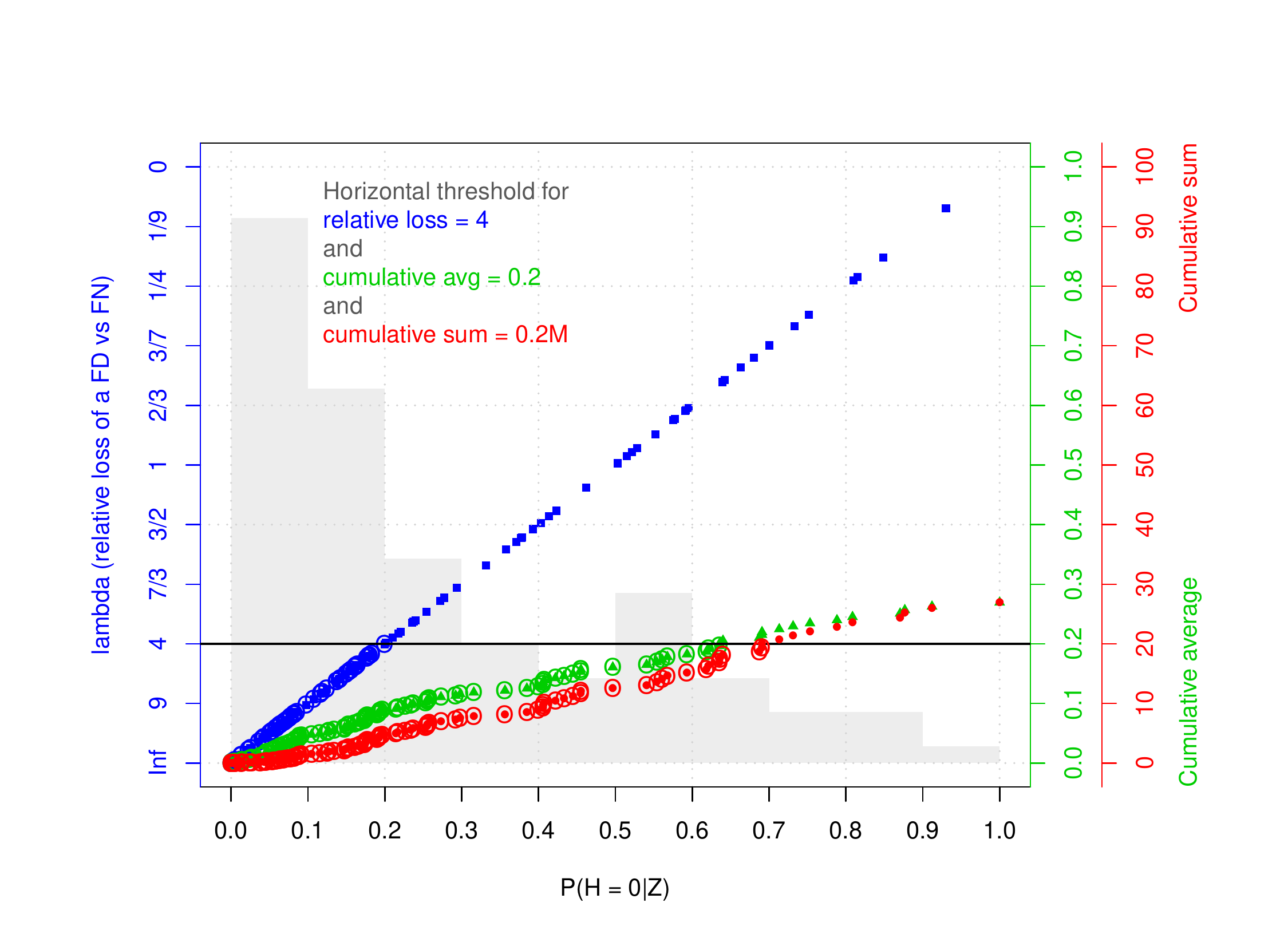}
\caption{A comparison of the various decision criteria, for $M=100$ artificially-generated posterior probabilities clustered around zero. The triangular points are plotted on the scale of $R_1$; the square points are plotted on the scale of $R_2$; the circular points are plotted on the scale of $R_3$. The horizontal threshold line illustrates the cutoff for all three decision criteria: $R_1$, where we want to make sure that fewer than 20\% of our discoveries are false; $R_2$ (which thresholds the raw probabilities), when we have specified a false discovery to be 4 times more costly than a false negative; and $R_3$, where we want to make sure that we have fewer than 20 total false discoveries.}
\label{compare_lossFcns_m0}
\end{center}
\end{figure}

\begin{figure}[!h]
\begin{center}
\includegraphics[trim={45 15 45 25mm}, clip, width = \textwidth]{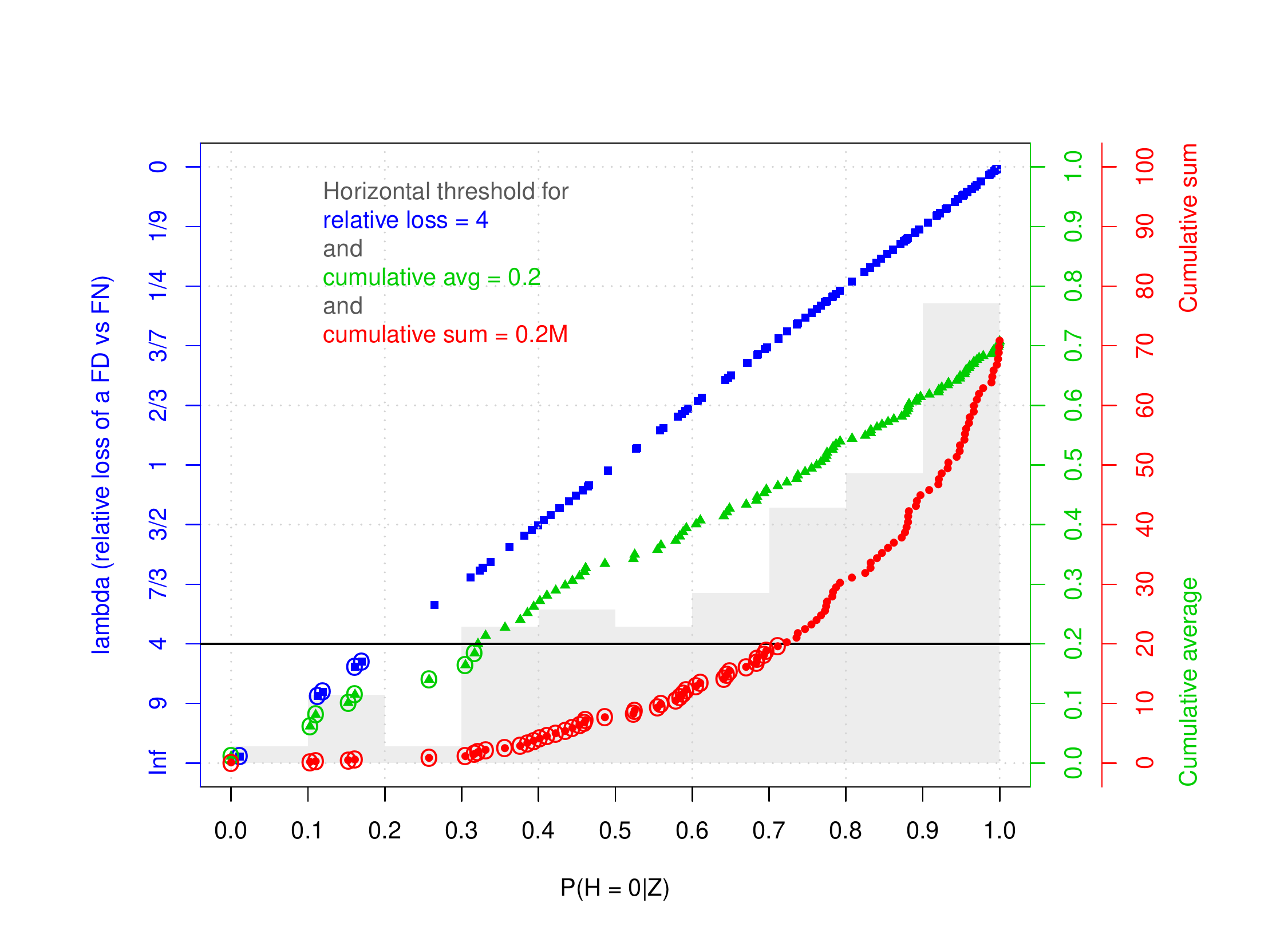}
\caption{A comparison of the various decision criteria, for $M=100$ artificially-generated posterior probabilities clustered around one. The triangular points are plotted on the scale of $R_1$; the square points are plotted on the scale of $R_2$; the circular points are plotted on the scale of $R_3$. The horizontal threshold line illustrates the cutoff for all three decision criteria: $R_1$, where we want to make sure that fewer than 20\% of our discoveries are false; $R_2$ (which thresholds the raw probabilities), when we have specified a false discovery to be 4 times more costly than a false negative; and $R_3$, where we want to make sure that we have fewer than 20 total false discoveries.}
\label{compare_lossFcns_m1}
\end{center}
\end{figure}

\begin{figure}[!h]
\begin{center}
\includegraphics[width = \textwidth]{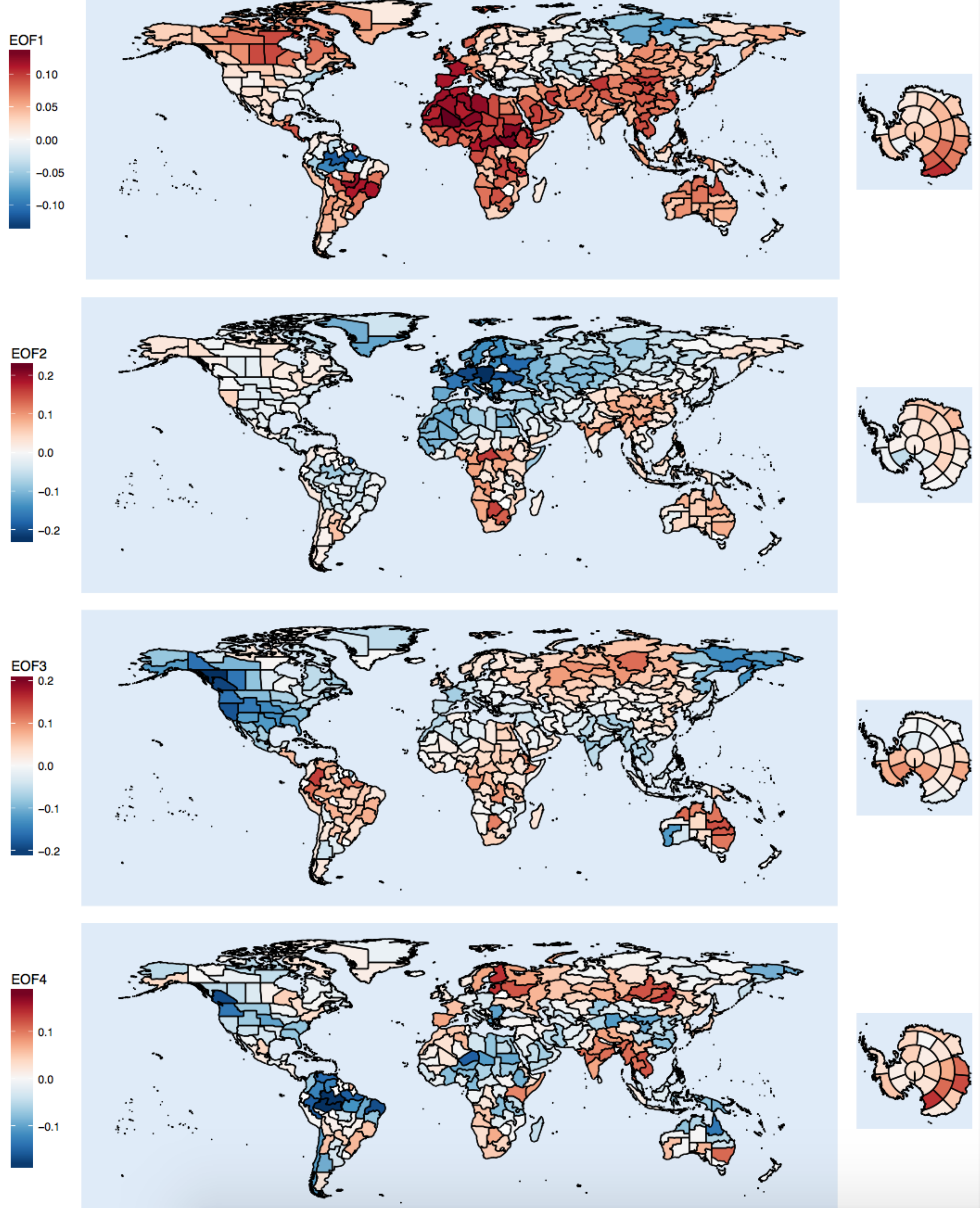}
\caption{The first four EOFs for the logit probability of a hot January over 1959-2014, for the factual scenario.}
\label{ALL_EOF_hotJan}
\end{center}
\end{figure}

\begin{figure}[!h]
\begin{center}
\includegraphics[width = \textwidth]{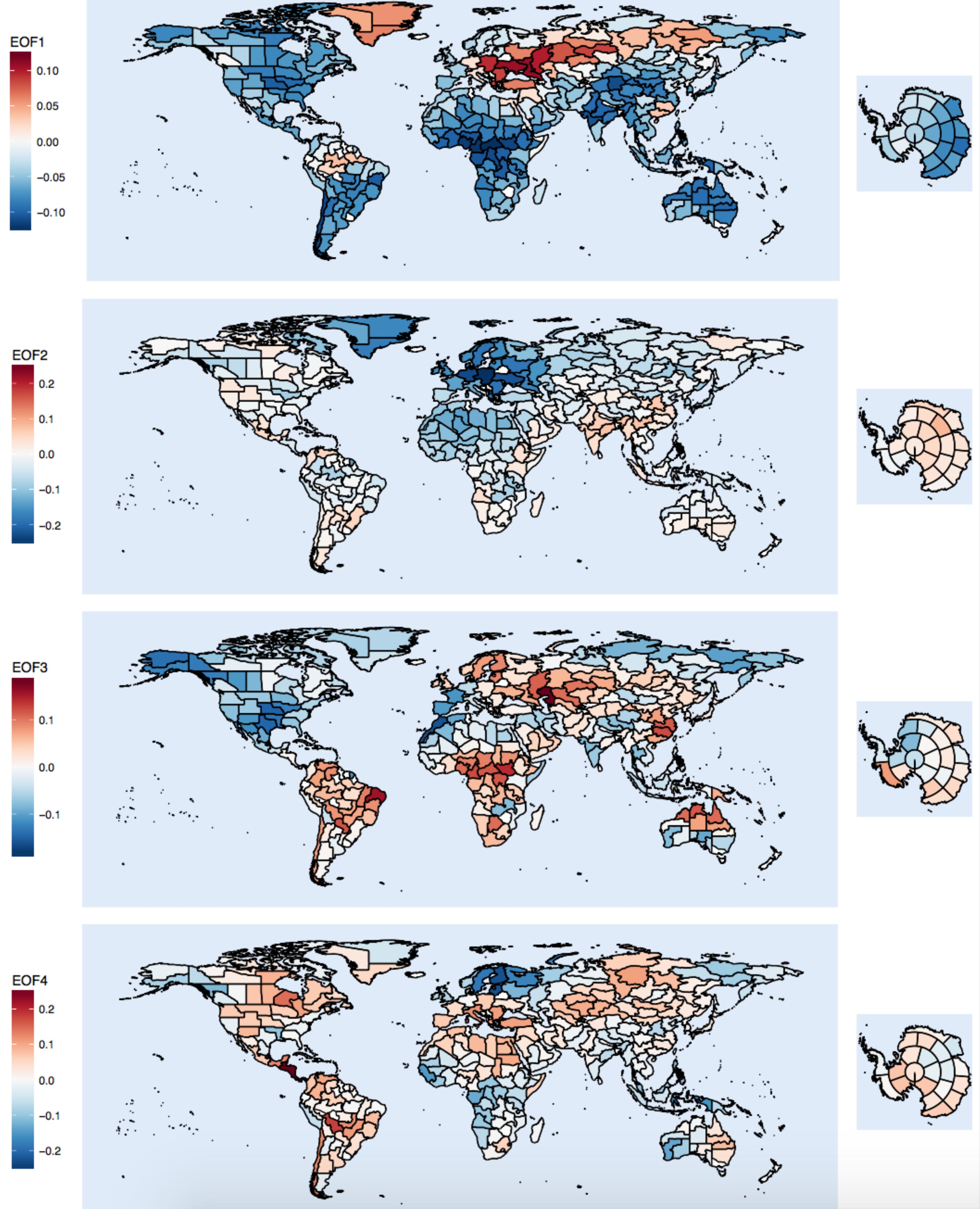}
\caption{The first four EOFs for the logit probability of a hot January over 1959-2014, for the counterfactual scenario.}
\label{NAT_EOF_hotJan}
\end{center}
\end{figure}

\begin{figure}[!h]
\begin{center}
\includegraphics[width = \textwidth]{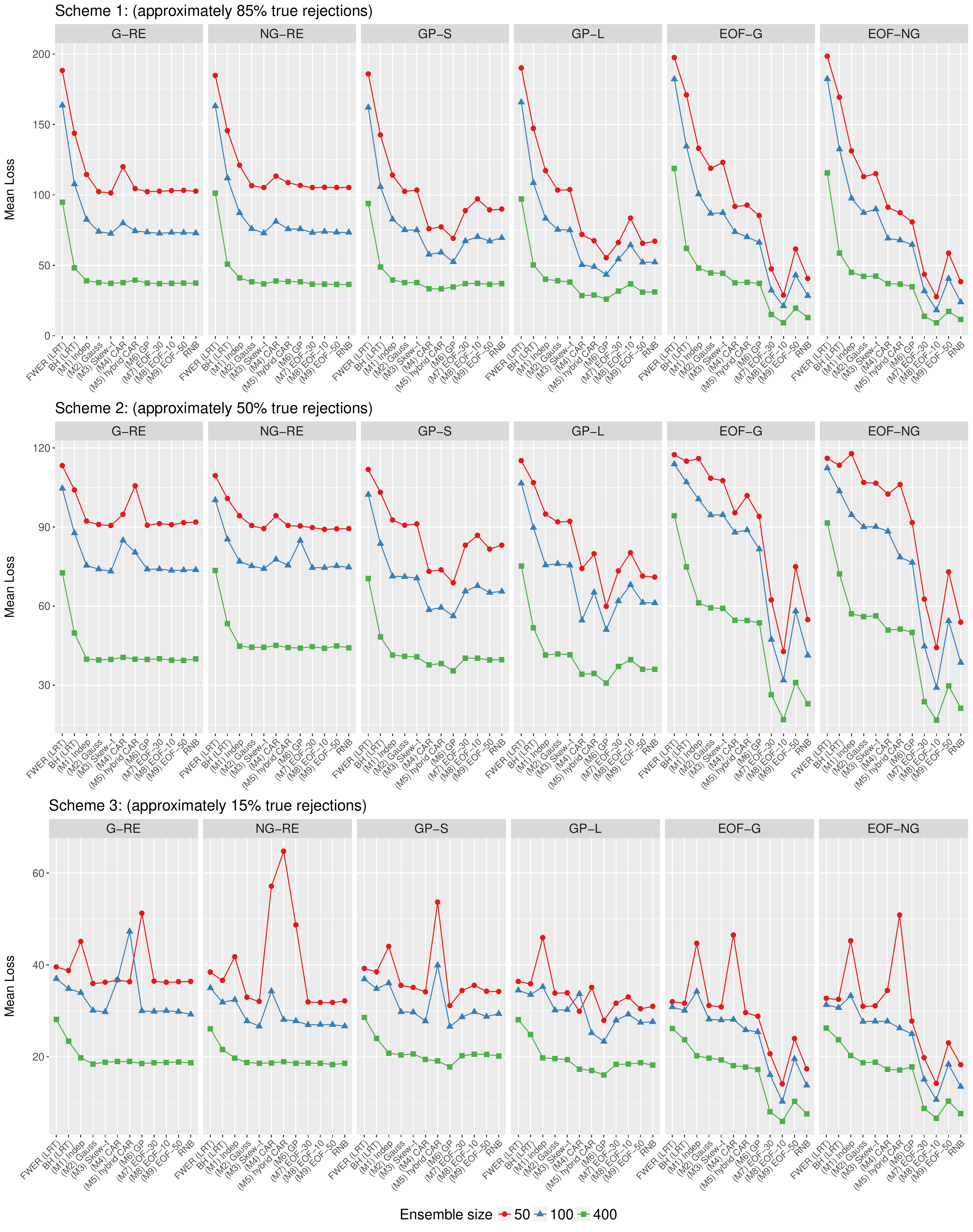}
\caption{Mean loss using the $R_2$ criteria, aggregated over the $N_{\text{rep}}=100$ replicates, for schemes 1, 2, and 3. Note that the $x$-axis in each subgrid corresponds to the different methods/fitted models.}
\label{L2_criteria}
\end{center}
\end{figure}

\begin{figure}[!h]
\begin{center}
\includegraphics[width = \textwidth]{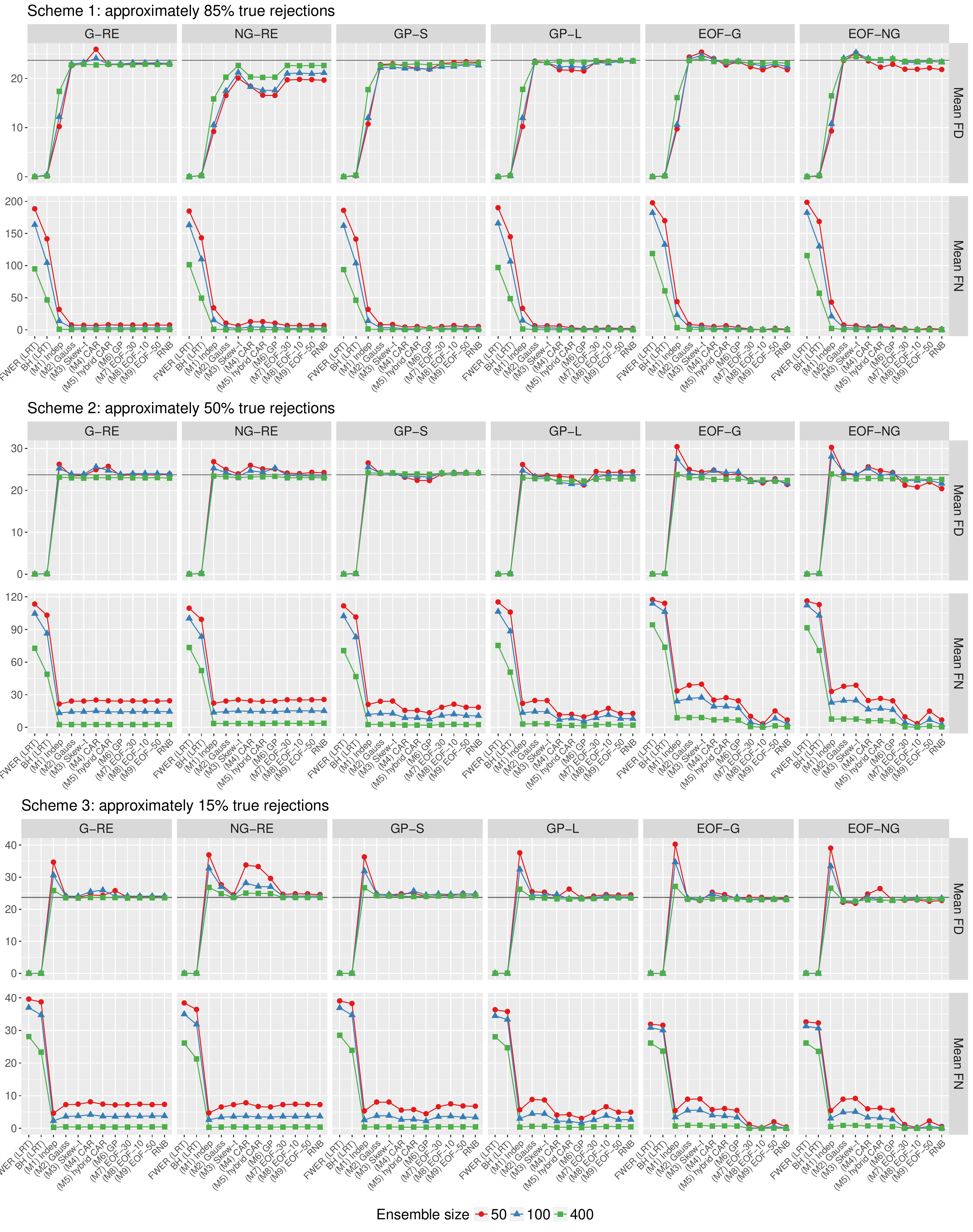}
\caption{Mean FD and FN using the $R_3$ criteria, aggregated over the $N_{\text{rep}}=100$ replicates, for schemes 1, 2, and 3. Note that the $x$-axis in each subgrid corresponds to the different methods/fitted models. The target of $\gamma=0.1M = 23.7$ is plotted for FD.}
\label{L3_criteria}
\end{center}
\end{figure}

\begin{figure}[!h]
\begin{center}
\includegraphics[width = 0.915\textwidth]{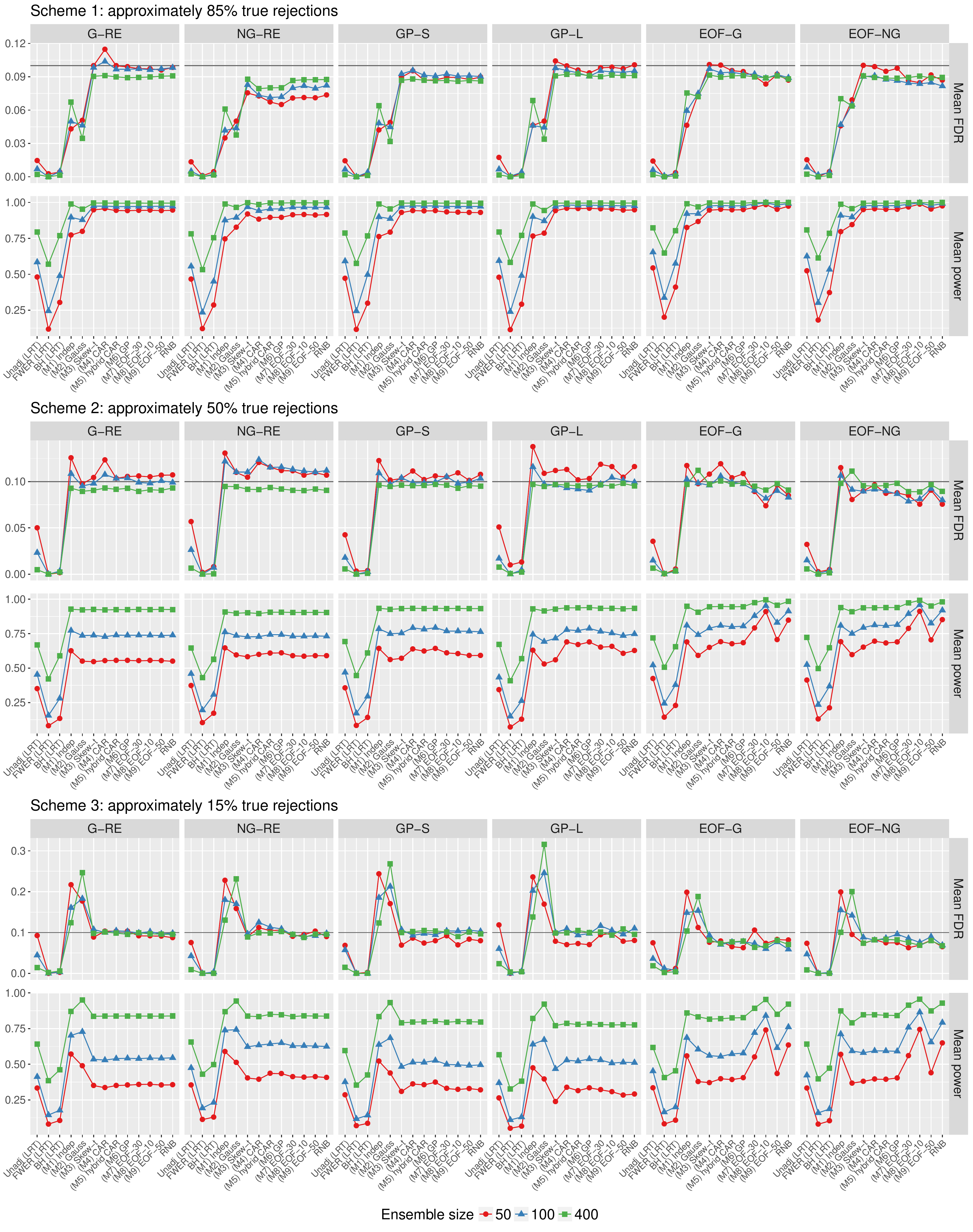}
\caption{Mean FDR and power using the $R_1$ criteria for the WRAF2 regions ($M=68$), aggregated over the $N_{\text{rep}}=100$ replicates, for schemes 1, 2, and 3. The target of $\alpha=0.1$ is plotted for FDR.}
\label{wraf2_L1_criteria}
\end{center}
\end{figure}

\begin{figure}[!h]
\begin{center}
\includegraphics[width = \textwidth]{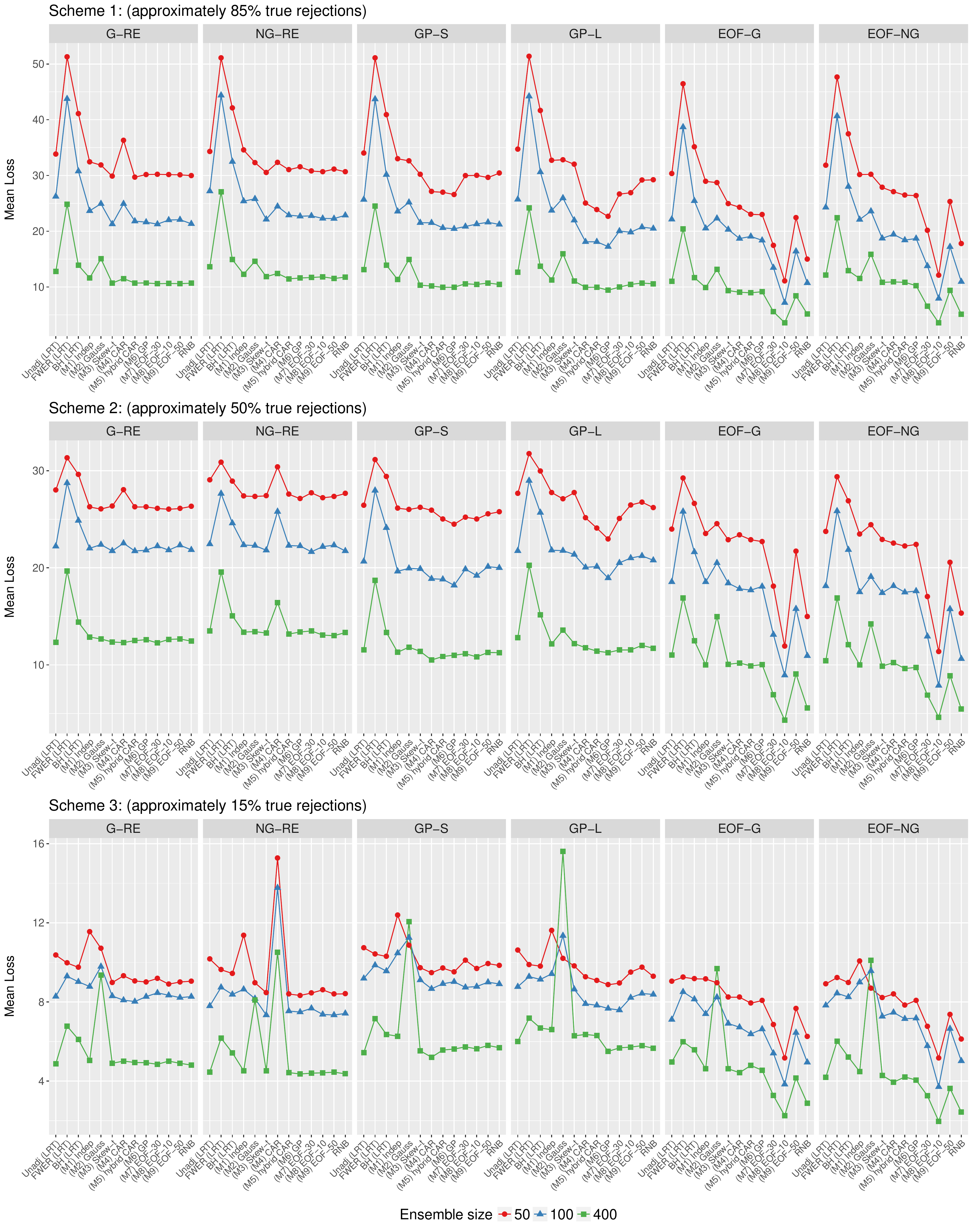}
\caption{Mean loss using the $R_2$ criteria for the WRAF2 regions ($M=68$), aggregated over the $N_{\text{rep}}=100$ replicates, for schemes 1, 2, and 3.}
\label{wraf2_L2_criteria}
\end{center}
\end{figure}

\begin{figure}[!h]
\begin{center}
\includegraphics[width = \textwidth]{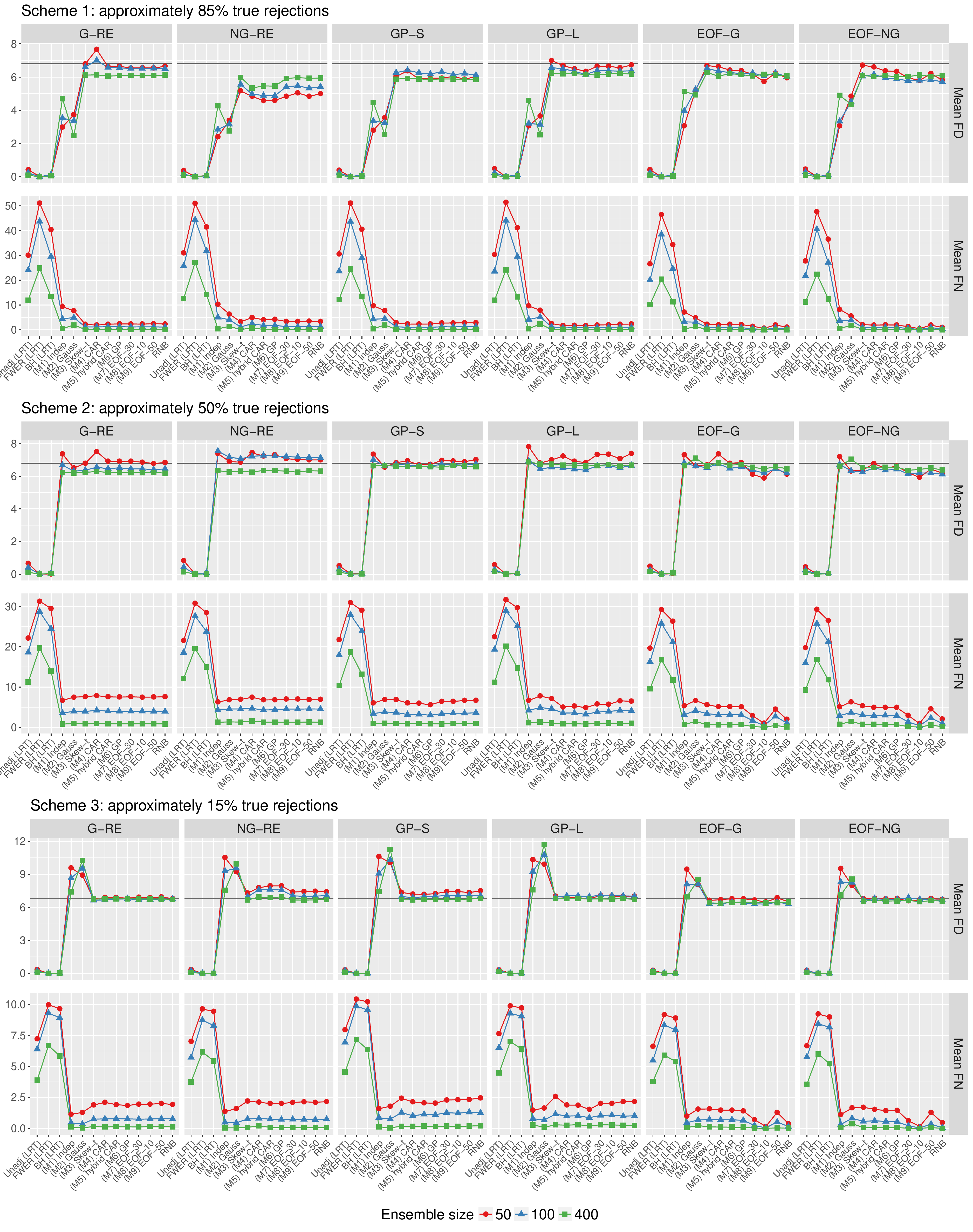}
\caption{Mean FD and FN using the $R_3$ criteria for the WRAF2 regions ($M=68$), aggregated over the $N_{\text{rep}}=100$ replicates, for schemes 1, 2, and 3. The target of $\gamma=0.1M = 6.8$ is plotted for FD.}
\label{wraf2_L3_criteria}
\end{center}
\end{figure}

Supplemental figures for the main text are shown in Figures \ref{compare_lossFcns_m0}, \ref{compare_lossFcns_m1}, \ref{ALL_EOF_hotJan}, \ref{NAT_EOF_hotJan}, \ref{L2_criteria}, and \ref{L3_criteria}. Results for the simulation study with the larger WRAF regions (WRAF2 with $M=68$) are shown in Figures  \ref{wraf2_L1_criteria}, \ref{wraf2_L2_criteria}, and \ref{wraf2_L3_criteria}.

\subsection{Main text}

\clearpage
\subsection{Results from simulation study with $M=68$ regions} \label{SuppFig_2}

%\begin{figure}[!h]
%\begin{center}
%\includegraphics[width = \textwidth]{FigureA7_wraf2.pdf}
%\caption{Larger WRAF2 land regions, with $M=68$ total regions.}
%\label{WRAF2}
%\end{center}
%\end{figure}
%\clearpage

\clearpage

% Skew-t details ===========================
\section{Centered parameterization for the skew-$t$ distribution} \label{skewt_param}

Note: the parameter symbols used in this section do not correspond to the symbols used in the main draft of the text.
\vskip2ex

\noindent \cite{Azzalini2003} introduced the skew-$t$ family of distributions, with probability density function
\begin{equation} \label{skewt}
f_{ST}(y; \xi, \omega, \alpha, \nu) = \frac{2}{\omega}t_\nu\left(\frac{y - \xi}{\omega}\right) T_{\nu+1}\left( \frac{\alpha(y - \xi)}{\omega} \sqrt{\frac{\nu+1}{\nu + (y - \xi)/\omega}} \hskip0.5ex \right),
\end{equation}
where $t_\nu$ and $T_\nu$ denote the probability density and cumulative distribution function, respectively, of a standard $t$ distribution with $\nu$ degrees of freedom. In (\ref{skewt}), $\xi \in \mathbbm{R}$ is a location parameter, $\omega \in \mathbbm{R}^+$ is a scale parameter, $\alpha \in \mathbbm{R}$ controls the skewness, and $\nu \in \mathbbm{R}^+$ controls the tail behavior. Unfortunately, as noted by \cite{ArellanoValle2008} (and others), using the ``direct'' parameterization $\theta_{D} = (\xi, \omega, \alpha, \nu)$ has both theoretical and practical problems: for example, the likelihood behaves strangely for a neighborhood of $\alpha = 0$, in that the profile likelihood for $\alpha$ has a stationary point at 0. Furthermore, at $\alpha = 0$, the expected Fisher information is singular, even though all of the parameters are identifiable. In practical terms, this means that the parameter estimates (especially $\xi$ and $\omega$) can trade off with one another to give qualitatively similar results for an individual data set.

To address this problem, \cite{ArellanoValle2008} discuss a ``centered'' parameterization (for the skew-normal distribution; a corresponding result holds for the skew-$t$), originally introduced by \cite{Azzalini2003}. Instead of $\theta_D$, the centered parameterization involves $\theta_C = (\mu, \sigma, \delta, \nu)$,
where
\[
\mu = \xi + \omega \sqrt{{2}/{\pi}} \frac{\alpha}{\sqrt{1+\alpha^2}}, \hskip3ex -\infty < \mu < \infty,
\]
\[
\sigma = \omega \sqrt{1 - \frac{2}{\pi} \frac{\alpha^2}{1+\alpha^2}}, \hskip3ex 0 < \sigma < \infty,
\]
and
\[
\delta = \frac{\alpha}{\sqrt{1+\alpha^2}}, \hskip3ex -1 < \delta < 1,
\]
with inverse transformations
\begin{equation} \label{skewt_trans}
\xi = \mu - \frac{\sigma}{\sqrt{1 - \frac{2}{\pi}\delta^2}}\sqrt{{2}/{\pi}}\delta, \hskip3ex
\omega = \frac{\sigma}{\sqrt{1 - \frac{2}{\pi}\delta^2}},\hskip3ex
\alpha = \frac{\delta}{\sqrt{1-\delta^2}}.
\end{equation}
(Note: $\nu$ is the same in both parameterizations.) Using $\theta_C$ avoids the problems associated with $\theta_D$; in practice, the likelihood associated with $\theta_C$ is given by (\ref{skewt}), after substituting in (\ref{skewt_trans}).

% Priors ===========================
\section{Prior specification for the parametric Bayesian models} \label{app_priors}

In general, the priors used for all parameters will be proper but diffuse, with fixed hyperparameters. The details for each model are as follows; all of the priors below are for both $k \in \{F, C\}$.

\vskip2ex
\noindent \textit{M1 Beta-binomial, independent across regions}

\noindent The only parameters in M1 are the probabilities themselves, which have already been assigned beta priors. The hyperparameters are set to $a_p = b_p = 1$, i.e., the probabilities are given an uniform prior. 

\vskip2ex
\noindent \textit{M2 Exchangeable Gaussian prior}

\noindent The parameters in M2 are the scenario-specific mean $\mu_k$ and variance $\tau^2_k$, with priors
\[
\mu_k \sim N(0, 10^2), \hskip4ex \tau_k \sim U(0, 100),
\]
where $N(a,b)$ is the Gaussian distribution with mean $a$ and variance $b$ and $U(c,d)$ is the uniform distribution on the interval $(c,d)$.

\vskip2ex
\noindent \textit{M3 Exchangeable skew-$t$ prior}

\noindent Following \cite{ArellanoValle2008}, M3 involves the scenario-specific ``centered'' parameters (see Appendix \ref{skewt_param}) location $\mu_k$, scale $\sigma_k$, skewness $\delta_k$, and degrees of freedom $\nu_k$. The prior distributions used are
\[
\mu_k \sim N(0, 10^2), \hskip2ex \sigma_k \sim U(0, 100), \hskip2ex \delta_k \sim U(-1, 1), \hskip2ex 1/\nu_k \sim U(0, 1).
\]

\vskip2ex
\noindent \textit{M4 CAR prior}

\noindent The parameters in M4 are the scenario-specific mean $\mu_k$ and variance $\tau^2_k$; however, because the CAR prior is improper, we fix $\mu_k = 0$ (see Appendix \ref{app_MCMC}). As before, $\tau_k \sim U(0,100)$.

\vskip2ex
\noindent \textit{M5 Hybrid CAR/exchangeable prior}

\noindent The parameters in M5 are the scenario-specific mean $\mu_k$, variance $\tau^2_k$, and mixture parameter $\lambda_k$, with priors
\[
\mu_k \sim N(0, 10^2), \hskip4ex \tau_k \sim U(0, 100), \hskip4ex \lambda_k \sim U(0,1).
\] 

\vskip2ex
\noindent \textit{M6 Gaussian process prior}

\noindent The parameters in M6 are the scenario-specific mean $\mu_k$, variance $\tau^2_k$, and spatial ``range'' parameter $\phi_k$, with priors
\[
\mu_k \sim N(0, 10^2), \hskip4ex \tau_k \sim U(0, 100), \hskip4ex \phi_k \sim U(0,c_\phi),
\] 
where $c_\phi = (1/2)\max\{ || \bfs_i -\bfs_j || \}$, since the range of the Gaussian process would not be expected to exceed one-half of the maximum distance between the region centroids. Note that the smoothness parameter for the Mat\'ern correlation function will be considered fixed, at $0.5$ (corresponding to an exponential correlation function).

\vskip2ex
\noindent \textit{M7/M8/M9 EOF-based structure with a Gaussian prior for a fixed number of coefficients}

\noindent The parameters in these three models are the scenario-specific mean $\mu_k$, EOF coefficients $\bfalpha_k$, scale $\sigma_k$, skewness $\delta_k$, and degrees of freedom $\nu_k$. As with the robust nonparametric Bayesian model, 
\[
\mu_k \sim N(0, 10^2), \hskip4ex \sigma_k \sim U(0, 100^2), \hskip4ex \delta_k \sim U(-1, 1), \hskip4ex 1/\nu_k \sim U(0,1).
\] 
In a more standard approach, the EOF coefficients (across $p=30$, $p=10$, and $p=50$) now have an exchangeable Gaussian prior:
\[
\alpha_{kl} \stackrel{\text{iid}}{\sim} N(0, \sigma^2_\alpha), \hskip2ex l = 1,\dots, p,
\]
where $\sigma_\alpha \sim U(0, 100)$.

% MCMC details ===========================
\section{Markov chain Monte Carlo} \label{app_MCMC}

The posterior distribution for each of the hierarchical models M2-M9 and RNB is not available in closed form, so we resort to Markov chain Monte Carlo (MCMC) methods to obtain samples from the joint posterior distribution for each model. All models are fit using the {\tt nimble} software for {\tt R} (\citealp{nimble}). While the MCMC is straightforward for RNB, M2, M3, M5, M6, M7, M8, and M9 (using standard Gibbs sampling with Metropolis Hastings steps), model M4 requires an adjustment to the standard MCMC (see the next section). The code used to fit these models are available in the online reproducibility documents.

\subsection{Computational details for the CAR parameterization}

Recall that computation for the CAR model is hindered by the fact that the intrinsic CAR prior is improper. This results in two problems: first, the random effects are identifiable only up to an additive constant; second, the CAR prior is undefined for the full random effects vector. While more sophisticated solutions to the first problem are possible, for the purposes of this work we simply set $\mu_k = 0$ to fix the identifiability problem.

\cite{RueHeld2005} outline steps to address the second problem. The CAR prior is
\[
p(\bfbeta_k | {\bf Q}_k, \tau^2_k) \propto \left| \tau^{-2}_k {\bf Q}\right|^{1/2} \exp\left\{ -\frac{1}{2} \bfbeta_k^\top  {\bf Q}_k \bfbeta_k \right\};
\]
however, the rank of ${\bf Q}$ is $M-1$ (${\bf 1}^\top{\bf Q} = 0$), so the determinant $\left| \tau^{-2}_k {\bf Q}\right|=0$. While the CAR prior is improper for an $M$-dimensional space, it is proper for a $(M-1)$-dimensional subspace. Following \cite{RueHeld2005}, the prior contribution to the posterior is actually 
\[
\widetilde{p}(\bfbeta_k | {\bf Q}_k, \tau^2_k) = (2\pi\tau^2_k)^{-\frac{(M-1)}{2}} \left(\prod_{i=1}^{M-1} \lambda_{ki} \right)^{1/2} \exp\left\{ -\frac{1}{2} \bfbeta_k^\top  {\bf Q}_k \bfbeta_k \right\},
\]
where $\{\lambda_{ki} : i = 1, \dots, M-1\}$ are the non-zero eigenvalues of ${\bf Q}_k$.

% Simulation study details ==============================
\section{Further details on the simulation study} \label{app_simstudy}

\subsection{Simulation scheme for each true state}

The six true states used as population distributions for the simulation study are listed in Table 1 of the main text. The actual sampling procedure for each true state is now outlined.

First, for the Gaussian random effects (G-RE), the logit probabilities are simply draws from a Gaussian distribution:
\[
\logit(p_k) \stackrel{\text{iid}}{\sim} N(m_k, v^2_k).
\]
Next, for the gamma random effects (NG-RE), the logit probability anomalies (i.e., deviations from the means $m_k$) are draws from a shifted gamma distribution:
\[
\logit(p_k) \stackrel{\text{iid}}{\sim} G(a_k, b_k) - c_k,
\]
where $a_k$ and $b_k$ are the shape and scale parameters, respectively. The Gaussian process samples (GP-S and GP-L) are first drawn collectively from
\[
\logit({\bf p}_k) \sim N_M(m_k{\bf 1}_M, {\bf S}),
\]
where the elements of ${\bf S}$ are $S_ij = v^2_k \mathcal{M}_{g_k}(||\bfs_i - \bfs_j||/r_k)$ (where $\mathcal{M}_g(\cdot)$ is the Mat\'ern correlation function and $\bfs_i$ is the centroid of region $i$) and then centered to have an empirical mean of zero.

It is slightly less straightforward to generate samples from EOF-G and EOF-NG, especially because the generated data needs to have properties comparable to the other simulations (in terms of the correct proportion of true rejections and empirical variance of the true log risk ratio). The following (somewhat complicated) scheme made this possible (the $k$ subscript has been omitted). 

\begin{enumerate}
\item For $j = 1, \dots, p$ (where we use $p = 30$ basis functions for the ``truth''), draw $\alpha_j \sim N(0, s^2_j)$.

\item Draw $x_j \stackrel{\text{iid}}{\sim} N(0, v^2)$ (for EOF-G) or $x_j \stackrel{\text{iid}}{\sim} k[G(b,c) - d]$ (for EOF-NG).

\item Calculate the probabilities as ${\bf p} = \logit^{-1}\big[m{\bf 1}_M + {\bf H}{\bf a} + {\bf x}\big]$.
\end{enumerate}

\subsection{Fixed hyperparameter values for the true states}

Tables \ref{GRE_hyp}-\ref{EOFNG_hyp} contain the fixed hyperparameters used to sample draws from the fixed population distributions across the $N_{\text{rep}}$ replicates. The values were determined after trial and error, and were set according to two criteria: first, that the true proportion of rejections would match up with the corresponding scheme, and second, that the variance of the true log risk ratio (empirically, over many replicates) would be approximately 0.9.

\begin{table}[!h]
\caption{Fixed hyperparameter values used for simulations from the Gaussian random effects (G-RE), across Schemes 1--3.}
\begin{center}
\begin{tabular}{|c|c|c|c|}
\hline
 			& Scheme 1 		& Scheme 2 		& Scheme 3 		\\ \hline\hline
$m_C$		& $\logit(0.08)$		& $\logit(0.08)$		& $\logit(0.08)$		\\ \hline
$m_F$		& $\logit(0.03)$		& $\logit(0.08)$		& $\logit(0.19)$	  	\\ \hline\hline
$v^2_C$, $v^2_F$  	& $0.72^2$		& $0.74^2$		& $0.775^2$		\\ \hline
\end{tabular}
\end{center}
\label{GRE_hyp}
\end{table}%

\begin{table}[!h]
\caption{Fixed hyperparameter values used for simulations from the shifted gamma random effects (NG-RE), across Schemes 1--3. Note: $a$ is the shape parameter and $b$ is the scale parameter.}
\begin{center}
\begin{tabular}{|c|c|c|c|}
\hline
 		& Scheme 1 		& Scheme 2 		& Scheme 3 		\\ \hline\hline
$m_C$	& $\logit(0.08)$		& $\logit(0.08)$		& $\logit(0.08)$		\\ \hline
$m_F$	& $\logit(0.03)$		& $\logit(0.08)$		& $\logit(0.18)$	  	\\ \hline\hline
$a_C$, $a_F$   & $4$		& $3.75$			& $3.5$			\\ \hline
$b_C$, $b_F$  	& $0.375$		& $0.4$			& $0.4286$		\\ \hline
$c_C$, $c_F$  	& $1.5$		& $1.5$			& $1.5$			\\ \hline
\end{tabular}
\end{center}
\label{NGRE_hyp}
\end{table}%

\begin{table}[!h]
\caption{Fixed hyperparameter values used for simulations from the spatial Gaussian process effects (GP-S and GP-L), across Schemes 1--3. Note: the distances in $\mathbbm{R}^3$ are re-scaled to have a maximum of 1 unit.}
\begin{center}
\begin{tabular}{|c|c|c|c|}
\hline
 		& Scheme 1 		& Scheme 2 		& Scheme 3 		\\ \hline\hline
$m_C$	& $\logit(0.08)$		& $\logit(0.08)$		& $\logit(0.08)$		\\ \hline
$m_F$	& $\logit(0.03)$		& $\logit(0.08)$		& $\logit(0.18)$	  	\\ \hline\hline
$v^2_C$, $v^2_F$   & $0.6$		& $0.6$			& $0.6$			\\ \hline
$r_C$, $r_F$ (short) & $0.06$		& $0.06$			& $0.06$		\\ \hline
$r_C$, $r_F$ (long) & $0.10$		& $0.10$			& $0.10$		\\ \hline
$g_C$, $g_F$  	& $2$		& $2$			& $2$			\\ \hline
\end{tabular}
\end{center}
\label{GP_hyp}
\end{table}%

\begin{table}[!h]
\caption{Fixed hyperparameter values used for simulations from the EOF effects with Gaussian discrepancy (EOF-G), across Schemes 1--3.}
\begin{center}
\begin{tabular}{|c|c|c|c|}
\hline
 					& Scheme 1 		& Scheme 2 		& Scheme 3 		\\ \hline\hline
$m_C$				& $\logit(0.08)$		& $\logit(0.08)$		& $\logit(0.08)$		\\ \hline
$m_F$				& $\logit(0.03)$		& $\logit(0.08)$		& $\logit(0.19)$	  	\\ \hline\hline
$s^2_j$, $j = 1, \dots 5$ 	& $3.5^2$			& $3.5^2$		 	& $3.5^2$			\\ \hline 
$s^2_j$, $j = 5, \dots 10$ 	 & $1^2$			& $1^2$			& $1^2$			\\ \hline 
$s^2_j$, $j = 10, \dots 30$ & $0.05^2$		& $0.05^2$		& $0.05^2$		\\ \hline \hline 
$v^2_C$, $v^2_F$   		& $0.01^2$		& $0.01^2$		& $0.01^2$		\\ \hline
\end{tabular}
\end{center}
\label{EOFG_hyp}
\end{table}%

\begin{table}[!h]
\caption{Fixed hyperparameter values used for simulations from the EOF effects with gamma discrepancy (EOF-NG), across Schemes 1--3.}
\begin{center}
\begin{tabular}{|c|c|c|c|}
\hline
 					& Scheme 1 		& Scheme 2 		& Scheme 3 		\\ \hline\hline
$m_C$				& $\logit(0.08)$		& $\logit(0.08)$		& $\logit(0.08)$		\\ \hline
$m_F$				& $\logit(0.03)$		& $\logit(0.08)$		& $\logit(0.19)$	  	\\ \hline\hline
$s^2_j$, $j = 1, \dots 5$ 	& $3.5^2$			& $3.5^2$		 	& $3.5^2$			\\ \hline 
$s^2_j$, $j = 5, \dots 10$ 	 & $1^2$			& $1^2$			& $1^2$			\\ \hline 
$s^2_j$, $j = 10, \dots 30$ & $0.05^2$		& $0.05^2$		& $0.05^2$		\\ \hline \hline 
$k_C$, $k_F$   		& $0.02$		& $0.02$			& $0.02$			\\ \hline
$b_C$, $b_F$   		& $5$		& $5$			& $5$			\\ \hline
$c_C$, $c_F$   		& $0.4$		& $0.4$			& $0.4$			\\ \hline
$d_C$, $d_F$   		& $2$		& $2$			& $2$			\\ \hline
\end{tabular}
\end{center}
\label{EOFNG_hyp}
\end{table}%

\end{appendix}

\end{document}